\documentclass[manuscript]{aastex}

\usepackage{graphicx}
\usepackage{float}
\usepackage[usenames, dvipsnames]{color}
\usepackage[normalem]{ulem}
\usepackage{natbib} 

\usepackage{color}           

\newcommand{\gui}[1]{{\bf {#1}}} 

\shorttitle{Topological Study of AR 12158}
\shortauthors{Zhao et al.}

\begin{document}

\title{Hooked flare ribbons and flux-rope related QSL footprints}

\author{Jie {Zhao}\altaffilmark{1}, {\bf Stuart A. {Gilchrist}}\altaffilmark{2,3}, Guillaume {Aulanier}\altaffilmark{2}, Brigitte {Schmieder}\altaffilmark{2}, Etienne {Pariat}\altaffilmark{2}, Hui {Li}\altaffilmark{1}}
\email{nj.lihui@pmo.ac.cn}
\altaffiltext{1}{Key Laboratory of Dark Matter and Space Astronomy, Purple Mountain Observatory, CAS, Nanjing 210008, China}
\altaffiltext{2}{LESIA, Observatoire de Paris, PSL Research University, CNRS,
                 Sorbonne Universit\'es, UPMC Univ. Paris 06, Univ. Paris-Diderot,
                 Sorbonne Paris Cit\'e, 5 place Jules Janssen, F-92195 Meudon, France}
\altaffiltext{3}{NorthWest Research Associates, 3380 Mitchell Lane, Boulder, CO 80301, USA}
\begin{abstract}
We studied the magnetic topology of active region 12158 on 2014
September 10 and compared it with the observations
before and early in the flare which begins
at 17:21 UT (SOL2014-09-10T17:45:00). Our results show that the
sigmoidal structure and flare ribbons of this active region observed by {\bf SDO/AIA}
can be well reproduced from a Grad-Rubin non linear force free field
extrapolation method. Various inverse-S and -J shaped magnetic field lines,
that surround a coronal flux rope,
coincide with the sigmoid as observed in different extreme ultraviolet wavelengths,
including its multi-threaded curved ends.
{\bf Also,}
the observed distribution of surface currents in the magnetic polarity where
it was not prescribed is well reproduced. This validates our
numerical implementation and set-up of the Grad-Rubin method.
The {\bf modeled} double inverse-J shaped Quasi-Separatrix Layer (QSL) footprints match
the observed flare ribbons during the rising phase of the flare,
including their hooked parts. The spiral-like shape of the latter
may be related to a complex pre-eruptive flux rope {\bf with} more than one turn of twist,
as obtained in the model.
These ribbon-associated  flux-rope QSL-footprints
are consistent with the new standard flare model in {\bf 3D},
with the presence of a hyperbolic flux tube located below an inverse {\bf tear drop shaped} coronal QSL.
This is a new step forward forecasting
the {\bf locations of reconnection and ribbons} in solar flares,
and the geometrical properties of eruptive flux ropes.
\end{abstract}


%
%

\section{Introduction}
\label{section_introduction}
Solar flares are the most violent examples of solar activity.
Eruptive flares accompany Coronal Mass Ejections (CMEs) and
are statistically the most energetic
type of flares \citep{2005JGRA..11012S05Y,2007ApJ...665.1428W}.
Although their pre-eruptive configuration and initiation mechanisms in three dimensions
have been studied for many years
(see e.g. the modeling of \citealt{2004A&A...425..345R} and \citealt{2014Natur.514..465A}, as well as the reviews of
\citealt{2014IAUS..300..184A,2015SoPh..tmp...63J,2015SoPh..tmp...64S}
with their accompanying references),
a standard model for eruptive flare has not yet been well established.
The old CSHKP model in two dimensions
\citep{1964NASSP..50..451C,1966Natur.211..695S,1974SoPh...34..323H,1976SoPh...50...85K}
remains an influential model that is often described as a ``standard'' model of eruptive flares.
In this model, the flare is powered by magnetic
reconnection at a vertical current sheet that forms in the
corona below a detached, upward-propagating plasmoid representing the CME.
Particles accelerated in the reconnection region travel along the
magnetic field to the Sun's surface. At the surface, the particles
heat the plasma causing the development of two bright flare ribbons at the footpoints
of magnetic separatrix field lines, which correspond to the reconnecting
field lines. In the CSHKP picture, the two ribbons are
straight lines that are parallel to the polarity inversion line
(PIL) and move away from each other as the flare proceeds.
Over the years, this general picture has been found to be consistent
with most observations, and has been extremely successful in explaining
the so-called two-ribbon flares. However, key elements that are
missing in the 2D CSHKP model are the shape, location, and dynamics
of the ribbons, most particularly their extremities,
and their link with the legs of the three dimensional (3D) erupting flux rope.

The issues with the CSHKP model were first addressed by a series of 3D
models that consider a non-force-free flux rope embedded in simple magnetic arcade
\citep{1996JGR...101.7631D}. In these models, the ribbons coincide with
the photospheric footprint of a Quasi-Separatrix Layer (QSL)
that encloses the flux rope.
The extremities of the
ribbons are hook shaped for weakly twisted flux ropes and are
spiral shaped for highly twisted flux ropes.

Later studies have shown that the QSLs of more realistic flux ropes
are also hook shaped. \citet{2007ApJ...660..863T} and \citet{2012A&A...541A..78P} analyze the analytic nonlinear
force-free-field (NLFFF) flux rope model of \citet{1999A&A...351..707T}.
We illustrate in Figure \ref{ep_fig_qsls} (adapted from \citealt{2012A&A...541A..78P}) the aforementioned
shapes for the hooks of the QSL footprints associated with flux ropes of various twists.
In this Figure,
panels (a) and (b) show the hook-shaped QSLs for flux ropes with
a moderate twist of 1 and 1.5 turns respectively.
Panel (c) shows the spiral-shaped QSLs for a highly twisted flux
rope with 2 turns. In all three cases the QSLs wrap around the
legs of the rope.
\citet{2011ApJ...738..167S} and \citet{2013A&A...555A..77J} analyze the
flux-cancellation magneto-hydrodynamical simulations of
\citet{2010ApJ...708..314A} and \citet{2012A&A...543A.110A} respectively.
\citet{Savcheva2012,2012ApJ...744...78S} analyze a NLFFF model of an observed
sigmoid. Their model is constructed using the flux-insertion magneto-frictional
method \citep{2004ApJ...612..519V}. Although the flux ropes are constructed
in different ways, J-shaped QSL-hooks are always present around
their intersection with the photosphere.

The idea that flare ribbons appear at the QSL footprints is also
supported by the identification of pairs of J-shaped
ribbons in a handful of observed eruptive flares that resemble the hooked QSL footprints
from the topological studies (e.g., \citealt[][]{2009SoPh..258...53C},
\citealt[][]{2011ApJ...738..167S}).

The results of the topological studies have inspired a new
model that extends the CSHKP model to 3D.
The model is presented in a series of papers
\citep{2012A&A...543A.110A,2013A&A...549A..66A,2013A&A...555A..77J}
and is discussed in the review of \citet{2015SoPh..tmp...63J}.
The model makes specific predictions regarding the geometry
of the pre-eruption magnetic field. In particular, it predicts
the presence of a particular magnetic structure called a
Hyperbolic Flux Tube (HFT) in the corona \citep{2002JGRA..107.1164T}.
The HFT is formed by intersecting QSLs and is the predicted site
of flare reconnection. The footprints of the HFT at the photosphere
are the locations of the flare ribbons, which are
hook shaped in the model.

This model is now being consolidated by new results.
As one example, \citet{Savcheva2012,2012ApJ...744...78S} find QSLs with a tear-drop geometry and
an HFT a few moments before the onset of the eruption of
a long-lived sigmoid, by using the flux-rope insertion method. As another example,
\citet{2014ApJ...787...88Z} find an HFT in a reconstruction of the coronal magnetic field
before a major flare on 2011 February 15 \citep{2011ApJ...738..167S,2012ApJ...748...77S}, by
using the Optimization method as first introduced by \citet{2000ApJ...540.1150W} and
further developed by \citet{2004SoPh..219...87W}.
In addition, again using the flux-rope insertion method, \citet{2015ApJ...810...96S} construct NLFFF models
for no less than seven sigmoids, with hooked-shaped QSL footprints,
that eventually developed into eruptive flares. The authors show, however,
that it is particularly difficult to reproduce the location of the hooked parts of the J-shaped
ribbons, and therefore to accurately model the flux rope endpoints \citep{Savcheva2016}.

These recent developments, however, may be questioned because the
models cited here all suffer from different limitations. The
Titov-D{\'e}moulin (hereafter TD) NLFFF model is analytical, symmetric and idealized.
The flux-cancellation MHD models are also idealized and their
flux-rope geometry depends on some observational-inspired (although still
parameterized) boundary driving. The Optimization-related
NLFFF comprises a low-altitude flux rope before the eruption, which is different
from the other NLFFF model \citep{2015ApJ...803...73I} for the same event.
The relative merits of the latter two approaches are
difficult to establish since the pre-eruptive sigmoid is poorly observed.
The flux-insertion method that was used to model a large number of sigmoids
does not use the observed vector magnetogram data as boundary conditions.
Instead, the method uses line-of-sight magnetogram data combined
with a parameterized insertion of axial and poloidal magnetic flux
in the corona. The magnitude of the inserted flux, and the shape
and length of the insertion region are free parameters of the method
that are adjusted by a trial-and-error procedure.
The trial-and-error procedure also includes
user-driven comparison of the models with coronal observations
(UV and X-rays images) to determine the best-suited model.


In this paper, we aim to test the association between QSLs
and flare ribbons (especially at the hook parts) by
modeling the coronal magnetic field of an eruptive sigmoid
using the Grad-Rubin method. This method uses the observed
vector magnetic field at the photosphere to construct the coronal
magnetic field using a nonlinear force-free model. Although there
are a number of methods for constructing NLFFF
(see reviews by \citealt{1989SoPh..120...19A,2008JGRA..113.3S02W,2012LRSP....9....5W,2013SoPh..288..481R}),
the Grad-Rubin method is the only one based on a well-posed boundary value problem.
This method has been able to reconstruct
twisted flux ropes and reproduce sigmoids as observed in X-rays \citep{2004A&A...425..345R,2009ApJ...693L..27C,2010ApJ...715.1566C}.
In addition, the method has been shown to be one of the best methods in terms of
satisfying the solenoidal condition for magnetic fields
(see \citealt{2009ApJ...696.1780D} for a comparison of methods).
Unlike other methods, the Grad-Rubin method is based on a well-posed
boundary value problem and does not unavoidably introduce a finite
divergence when observed boundary conditions are imposed.
In this paper, the NLFFF calculation is done in spherical geometry using
the recently developed Current-Field Iteration in Spherical Coordinates
(CFITS) code \citep{2014SoPh..289.1153G}.

We consider the sigmoid that was observed in extreme ultraviolet (EUV) with the
{\it Solar Dynamics Observatory } \citep[\textit{SDO;}][]{Pesnell2012},
within the NOAA AR 12158, on 2014 September 10. Our choice for this event is motivated
by the recent report of so-called slipping reconnection during the
eruption, as evidenced by loop displacements along the J-shaped
ends of the flare ribbons \citep{2015ApJ...804L...8L}. This slipping behavior
corresponds to the regime under which reconnection takes place in
QSLs in general \citep{2006SoPh..238..347A} and in MHD simulations of
erupting sigmoids in particular \citep{2013A&A...555A..77J}.
This event occurs within a bipolar magnetic environment that
is simpler than that of the few other similar events previously
reported \citep{2014ApJ...791L..13L}.

Our study has two objectives. The first is to test if
the Grad-Rubin method in general, and the CFITS code in particular,
can reproduce all the following observational features: the sigmoid,
the flare ribbons and their J-shaped extremities, and the
photospheric current distribution, especially within the polarity where the
observed currents are not prescribed as boundary conditions (see
discussion in \citealt{2009ApJ...700L..88W}). The second is to
test the robustness of earlier findings, and  to assess whether
the 3D model proposed by \citet{2015SoPh..tmp...63J}
can be qualified as standard or not.

The paper is organized as follows: In Section \ref{section_observations}
we present the observations of the sigmoid of AR 12158.
In Section \ref{section_extrapolation} we present the results of the
NLFFF extrapolation. In Section \ref{section_topology} we compare the magnetic
topology obtained from the extrapolation and the flare ribbons.
In Section \ref{section_discussion} we discuss the results.
In Section \ref{section_conclusion} we give a summary and conclusion.

%
%
%
%
%
%
\section{Observations}
\label{section_observations}

NOAA AR 12158 rotated onto the disk around 2014 September 3.
In the days leading up to the eruption on September 10,
the region was bipolar and consisted of a single large
spot surrounded by plage. Panel (a) of figure \ref{sg_fig1}
shows the radial magnetic field of the region derived from an
Helioseismic and Magnetic Imager (HMI) magnetogram \citep{2012SoPh..275..207S}
taken on September 10 at 15:24 UT (about two hours before the eruption).
The image shows that the central spot is positive and is flanked by a diffuse region
of negative field to the south. Panel (b) shows the radial
electric current density $J_r$ for the region derived from the
curl of the HMI vector magnetic field at the same time.
A concentrated arc of negative $J_r$ dominates the positive spot (shown by the arrow $A1$).
The current over the negative polarity is generally diffuse, with
a single strong concentration to the east of
the main spot (indicated by the arrow $C1$).

The region was observed by the {\it Atmospheric Imaging Assembly} (AIA)
onboard {\it SDO} \citep[][]{2012SoPh..275...17L} and in soft X-rays (SXR)
with the {\it X-Ray Telescope} \citep[XRT;][]{2007SoPh..243...63G} instrument on Hinode \citep{2007SoPh..243....3K}.
We are primarily interested in the reconstruction of  the sigmoid on September 10. There it is primarily
visible in the three specific EUV channels at 94 \AA, 131 \AA, and 335 \AA (see
top panels of Figure \ref{sg_fig0}) as well as in SXR. We do not show the latter in this paper because,
for this particular sigmoid, the features
that are visible in SXR are very similar to those observed in 94 \AA (where the sigmoid is also more prominent
than in all the other EUV channels of AIA.)
The sigmoid has an inverse-S shape and it displays several individual loops and substructures.

Numerical simulations of sigmoid formation
and eruption \citep{2014Natur.514..465A}, of kink \citep{2004A&A...413L..23K} and
torus \citep{2010ApJ...708..314A} instabilities of solar-like flux ropes,
and of the distribution of their electric currents \citep{2014ApJ...782L..10T,2015ApJ...810...17D}
all {\bf together} suggest that sigmoids are formed by the ensemble
of the double-J shaped arcades in sheared magnetic fields,
and that when those arcades reconnect, they turn into fully S-shaped loops which progressively
build-up a flux rope. In the frame of these models,
the sigmoid has the same handedness with the flux
rope, {\bf i.e.} the inverse-S shape is associated with a left-handed twist and a negative helicity,
and the electric current density $\mathbf J$ in the center of the flux rope is anti-parallel
to the {\bf magnetic} field $\mathbf B$.
The distribution of $J_r$ and $B_r$
as {\bf observed} near the sigmoid ends ({\bf the regions} that primarily {\bf display} $J_r~B_r\le 0$, see
Figure \ref{sg_fig1}) is consistent with this scenario.

On 2014 September 10, the sigmoid erupted producing an
X1.6 class flare and a CME\footnote{\url{http://www.swpc.noaa.gov}}.
Based on the GOES  1$-$8 \AA\ soft X-ray light curve, the flare begins at
17:21 UT, reaches its peak at 17:45 UT, and ends at 18:20 UT.
This event is the focus of several observational studies:
\citet{2015ApJ...804...82C} performs a spectroscopic study of the
sigmoid and the flare ribbons, while \citet{2015ApJ...804L...8L} reports
the observation of slip-running reconnection
that \citet{dud2015} investigate in more detail.
The Solar Object Locator (SOL) for this event is SOL2014-09-10T17:45:00.

%
%
%
%
%
%

\section{Nonlinear Force-free extrapolation of AR 12158}
\label{section_extrapolation}

We construct a NLFFF model of the coronal
magnetic field of AR 12158 using the Current-Field
Iteration in Spherical Coordinates code (CFITS)
\citep{2014SoPh..289.1153G}. The code is a numerical  implementation of the
Grad-Rubin method \citep{24756} in spherical coordinates different from the one developed by
\citet{2013A&A...553A..43A}.
The present study is the first time the CFITS code has been
applied to solar data rather than test cases \citep{2014SoPh..289.1153G}.

NLFFF modeling of the corona is the subject
of multiple reviews \citep{2008JGRA..113.3S02W,2012LRSP....9....5W},
so we only briefly outline the general approach here. The basic
assumption is that the magnetic field in the corona, $\mathbf B$,  exists in a force-free state,
i.e.\ $\mathbf J \times \mathbf B = 0$, where $\mathbf J$ is the
electric current density. This condition combined with
Amp{\`e}re's law provides a set of partial-differential equations (the force-free equations)
for the magnetic field in the volume \citep{9559}.
The equations can be solved subject to boundary
conditions on the magnetic field and electric current at the photosphere
and other boundaries. The force-free equations are generally nonlinear,
and specialized numerical methods are required to solve them
\citep{2012LRSP....9....5W}.

In spherical geometry, the boundary conditions are imposed on a
spherical shell of a given radius. We impose boundary conditions
on a shell located at $R_{\sun}$. In practice, the boundary conditions
are derived from vector magnetogram data. We note that {\bf these} data
{\bf represent} the magnetic field around the height of line formation,
which {\bf do} not necessarily correspond to $R_{\sun}$, but we make
the approximation that {\bf the two coincide}. The specific boundary
conditions imposed are the normal component of $\mathbf B$,
\begin{equation}
  B_r |_{r = R_\sun},
  \label{bcs_ff1}
\end{equation}
and the distribution of the force-free parameter,
$\alpha = \mu_0 \mathbf J \cdot \mathbf B/B^2$, which
we take as
\begin{equation}
  \alpha_0 = \mu_0  \left. \frac{ J_r }{B_r} \right |_{r=R_\sun},
  \label{bcs_ff2}
\end{equation}
because vector-magnetogram data does not provide the vector-current density $\mathbf J$.
In fact, to prescribe $\alpha_0$ everywhere in the boundary
is an over specification of the boundary value problem \citep{24756}.
Instead, $\alpha_0$ is only prescribed over one polarity of $B_r$,
i.e.\ we fix $\alpha_0$ either at points in the boundary where
$B_r>0$ or at points in the boundary where $B_r<0$.
We refer to the two choices as the $P$ and $N$ boundary
conditions respectively.

We derive the $P$ and $N$ boundary conditions from vector-magnetogram
data from {\it HMI/SDO}. Specifically, we
use data taken at 15:24 UT on September 10 from the
720s Space-Weather HMI Active Region Patch (SHARP)
series \citep{2014SoPh..289.3549B} with HARP number 4536.
We determine $J_r$ by computing the curl of the magnetic field
at the photosphere after first smoothing the transverse components
using a Laplace scheme \citep{Press:2007:NRE:1403886}.
The smoothing reduces strong gradients in $\alpha_0$,
which are difficult to resolve numerically in the modeling.
In addition,
we set $\alpha_0=0$ where $|B_r|<50\,{\rm G}$,
because the transverse magnetic field is typically
poorly measured in weak field regions.

Panels (a)-(c) of Figure \ref{sg_fig1} shows the distributions of $B_r$ and $J_r$
derived from the HMI data. The data are saturated at the limits
indicated by the color bars. Panel (a) shows $B_r$.
Panel (b) shows $J_r$ without any smoothing applied.
Panel (c) shows $J_r$ computed after first smoothing the transverse components of
the magnetic field. Although the smoothing removes a lot of fine
structure, the overall distribution of $J_r$ is left intact.
The region is viewed as seen from Earth, and the yellow curve
demarcates the extent of the HMI magnetogram at the photosphere.
The green and purple lines are the $+200\,{\rm G}$ and $-200\,{\rm G}$
contours of $B_r$ respectively.

Unlike in the corona, the magnetic field at the photosphere
is not force free \citep{1995ApJ...439..474M}, which
raises concerns about the NLFFF modeling (e.g.\ \citealt{2009ApJ...696.1780D}).
This problem is generally treated by pre-processing
the vector magnetogram data to minimize forces in the boundary
\citep{2006SoPh..233..215W}. Other methods modify the photospheric boundary data in both polarities
during the solution of the force-free equations themselves \citep{2009ApJ...700L..88W}.
In either case, the magnitude of $\alpha$ in the boundary is
typically reduced. We adopt an alternative
approach, and reduce $\alpha$ in the boundary by a
parameterization of the boundary conditions.
Our parameterization is crude -- we simply scale
$\alpha_0$ by a constant factor $k$.
We do not parameterize $B_r$.
The parameterization is chosen to model the decrease in $\alpha$ from
the photosphere to the corona \citep{2014ApJ...793...15G},
although it does not model changes
to the spatial distribution of $\alpha_0$, which are likely important.

In the following, we will refer to the original distribution of $\alpha$ derived
from the HMI data before smoothing and scaling as $\alpha^{\rm raw}_0$,
and we refer to the distribution of $\alpha$ after smoothing and scaling
as $\alpha_0$. The latter is used as boundary conditions for the
modeling.

For our calculation, we use a spherical-polar coordinate mesh with
$256$ points in the radial direction, $273$ points in latitude, and $304$ points
in longitude. The grid has a uniform spacing
of {\bf $0.7 {\rm Mm}$} in both latitude and longitude, and a uniform spacing of
$1.1\,{\rm Mm}$ in the radial direction. The grid extents radially
to a distance of $0.4$ solar radii ($\approx268\,{\rm Mm}$) above the photosphere.

We apply CFITS to the $P$ and $N$ boundary conditions.
We find that the $N$ solution is close to a potential field ---
the magnetic energy and the field lines are close to the initial
potential field. The $N$ solution does not contain a flux rope,
regardless of the degree of smoothing or the choice of the $k$ parameter.

The boundary conditions for the $N$
solution are derived from the HMI observations in a
dispersed plage region.
There the sizes of network flux-concentrations, as observed at the
high-resolution of HMI, are not very large as compared with the pixel
size of the instrument. The associated electric
current densities, as calculated from Amp\`ere's law, are therefore
very fragmented, with little coherent structure
on average (even though some current concentrations are still noticeable).
As a consequence, the NLFFF extrapolation either becomes intractable, when one
wishes to keep such spiky current densities as boundary conditions, or results in a quasi-potential
field with only of few regions that display moderately sheared-arcades, when
the current densities are smoothed enough to make the calculation feasible.
As a result, we discard the $N$ solution
and only present the $P$ solution here. The $P$ solution only relies
on $\alpha_0$ values in the positive polarity and so is unaffected
by the noise in the negative polarity --- this is an advantage of the
Grad-Rubin method over methods that prescribe boundary conditions over the
entire boundary.

For the $P$ boundary conditions,
we perform multiple calculations with different values of $k$ in
the range $[0.5-1.0]$ and different degrees of smoothing. We
have chosen the solution that best matches the aforementioned AIA data to present
here. We note that the convergence of the method  for $k=1$ is poor.
This can occur in the presence of large electric currents, and has been
reported for other Grad-Rubin codes \citep{2012SoPh..276..133G}.
The convergence of the method is generally improved for smaller $k$,
where the solution becomes closer to a potential field, although the
resulting field does not closely match the AIA image
(we do not present comparisons between these solutions and the AIA images).

Figure \ref{sg_fig0} (d) - (h) {\bf show} the field lines of the NLFFF model
after $30$ Grad-Rubin iterations applied to the $P$ boundary
conditions with $k=0.66$. Panels (d)-(f) show the field lines
of the solution superimposed on AIA images of the region. The
red, yellow, and orange lines show the inverse-S and -J shaped magnetic field lines which appear to trace the sigmoid.
The purple field lines show a flux rope. But it has no observational counterpart,
{\bf neither in the EUV AIA channels (94 \AA, 131 \AA, 335 \AA)}
, nor in the XRT images in SXR. Panels (g) and (h) show three dimensional
views of the same sets of field lines in addition to a set of
arcade field lines in blue.

For the $P$ solution, we impose boundary conditions on $\alpha$
at points in the photosphere where $B_r>0$, while the value of $\alpha$ at points
in the photosphere where $B_r <0$ is determine by solving the force-free model.
We refer to the distribution of $\alpha$ at the photosphere obtained
from the modeling as $\alpha_0^{P}$.  Generally, where $B_r>0$,
$\alpha_0^{P} = \alpha_0$, except in open field regions where
the code sets $\alpha_0^{P}=0$ \citep{2014SoPh..289.1153G}. A
priori we do not expect $\alpha_0^{P} = \alpha_0$ where $B_r<0$, because
the distribution $\alpha_0$ is not consistent with a force-free model,
i.e. the model will only {\bf very} approximately recover the magnetogram
currents. In addition, since
$\alpha_0$ does not match $\alpha_0^{\rm raw}$ due to smoothing
and scaling, a priori we do not expect a close match between
$\alpha_0^{P}$ and $\alpha_0^{\rm raw}$.
However the observed distribution of currents over the negative polarities
matches well with the reconstructed distribution
of currents. This validates our present
numerical implementation of the Grad-Rubin method.
Panel (d) of Figure \ref{sg_fig1} shows the radial current
density, $J_r$, constructed from $\alpha_0^{P}$. This can be compared
with the distributions of $J_r$ in panel (c) which
is constructed from $\alpha_0$. There are noticeable differences between
the two distributions, nevertheless, the model recovers the
strong current concentration at $(X,Y)\approx (-150",150")$. This
feature is also present in the unsmoothed data, as indicated by
the arrow $C1$ in panel (b). The good correspondence between the modeled
$\alpha$ distribution and the observed $\alpha$ distribution
suggests our model field is a plausible model of the coronal magnetic
field of AR 12158.

To measure the divergence of our solution we compute the metric
$\langle |f_i| \rangle$ from \citet{2000ApJ...540.1150W}. This
metric measures the flux imbalance over each grid cell averaged
over the whole domain. Specifically,
\begin{equation}
  f_i = \frac{\int \mathbf B \cdot \mathbf{dS}}{\int |\mathbf B \cdot \mathbf{dS}|}
\end{equation}
where the integrals are over the boundaries of each spherical grid
cell, and $\langle . \rangle$ denotes the average over all cells
in the domain. For our solution we find
$\langle |f_i| \rangle \approx 6.5\times 10^{-4}$. This
value, obtained for a highly-stressed magnetic configuration,
is comparable to those found in \citet{2015ApJ...811..107D} for
force-free extrapolations {\bf in} Cartesian coordinates using a variety of
methods applied to a much less stressed solar active region (some of
which still being able to perform relatively better or worse than
this value).

%
%
%
%
%
%
\section{Topological analysis and observational comparison}
\label{section_topology}

We compute the QSLs of the force-free model obtained from
one HMI magnetogram from 15:24 UT for comparison with the observed
flare ribbons during the initial rising phase of the flare.

%
\subsection{Topology and QSLs}
\label{section_q_maps}

With the aim of determining the QSLs, we compute the squashing factor
$Q$ for the nonlinear force-free
model using the formula of \citet{2012A&A...541A..78P}.

The squashing factor is a measure of the gradient in the
field line connectivity \citep{2002JGRA..107.1164T}, and
thin layers where $Q$ is ``large'' are defined as QSLs.
In principle, QSLs are three dimensional.
We have chosen two-dimensional slices for comparison with observations and models.
For comparison with observations, only the projections of the QSLs at the photosphere
are relevant for comparison with the flare ribbons. For comparison with the models,
we have computed QSLs on a slice through the flux rope in order to identify the Hyperbolic Flux Tube (HFT)
and compare with similar QSL calculations made with model flux ropes. This kind of comparison
has already been achieved in earlier studies of NLFFF models of sigmoids
\citep{Savcheva2012,2015ApJ...810...96S,Savcheva2016,2014ApJ...787...88Z}.
QSLs have rarely been computed in three dimensions \citep[see e.g.][]{2015ApJ...806..171Y},
and to the authors' knowledge this has been achieved for flux
rope models only in the case of one analytical
TD rope \citep{2007ApJ...660..863T}.

The NLFFF extrapolation is performed
in spherical geometry, however we lack the numerical libraries
for computing $Q$ in this geometry. Since the region is
compact, we neglect the curvature of the photosphere and compute $Q$
using existing software designed for Cartesian geometry \citep[as in][]{Savcheva2012}.

Furthermore, for the $Q$ calculation, we interpolate the
extrapolation magnetic field onto a grid that is uniformly spaced
in all three Cartesian dimensions. The grid size is $\approx$ 0.36 Mm.
The reference boundary for the $Q$ computation, which defines the footpoint heights of
each integrated field line, is chosen at $Z\approx 1.5\, {\rm Mm}$ to highlight
the flux-rope related QSLs, and to exclude the QSLs related to small-scale
polarities at the photosphere \citep[cf.][]{Savcheva2012}. In the
following we refer to the plane at $Z\approx 1.5\, {\rm Mm}$ as the
lower boundary. We compute the distribution of $Q$ at the lower boundary and a
vertical plane that cuts the flux rope.

\begin{figure}
\centering

  \includegraphics[scale=0.52, angle=0, trim = 0.5cm 0.0cm 1.0cm 0.0cm, clip]{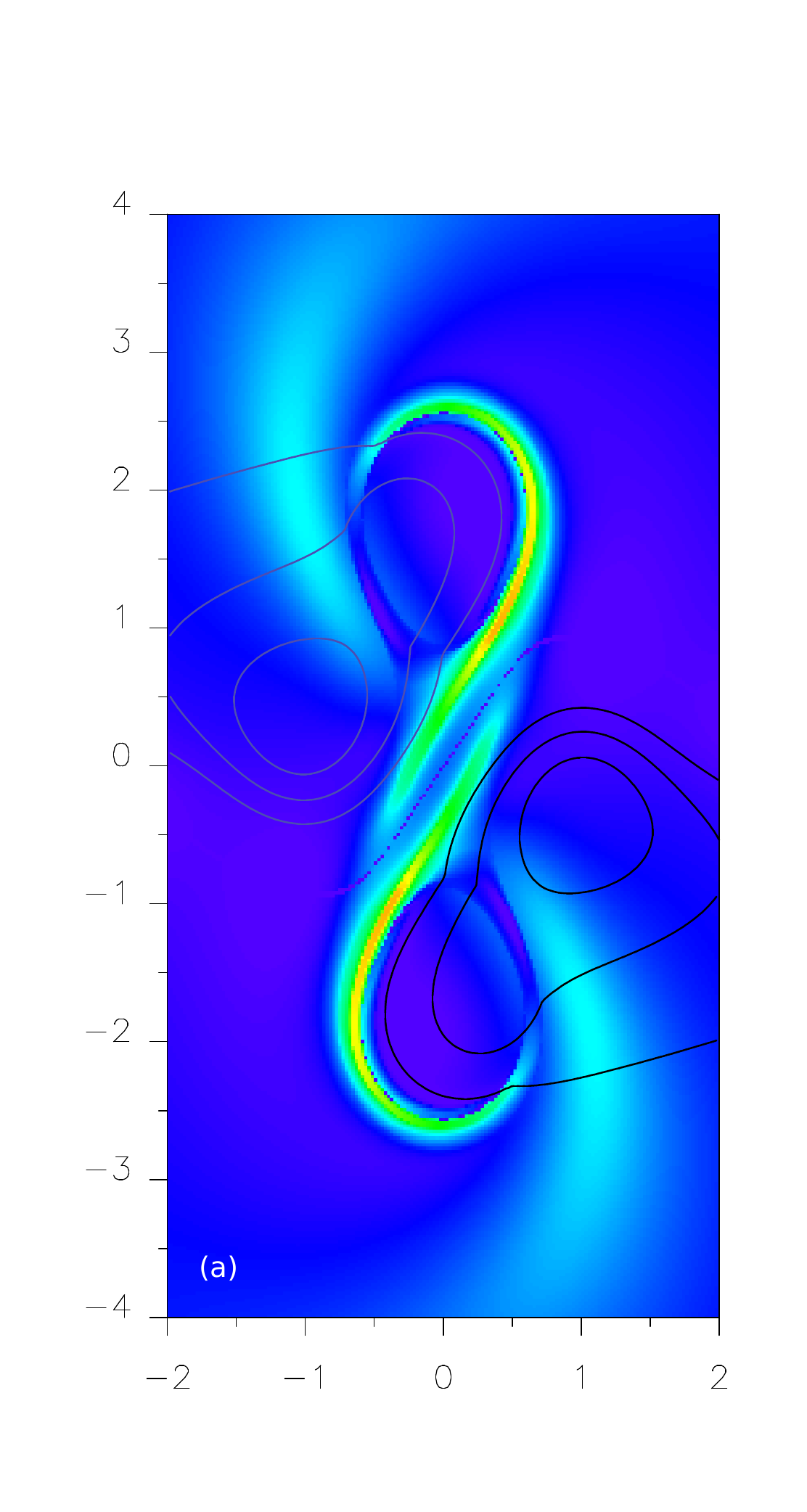} %
  \includegraphics[scale=0.52, angle=0, trim = 0.5cm 0.0cm 1.0cm 0.0cm, clip]{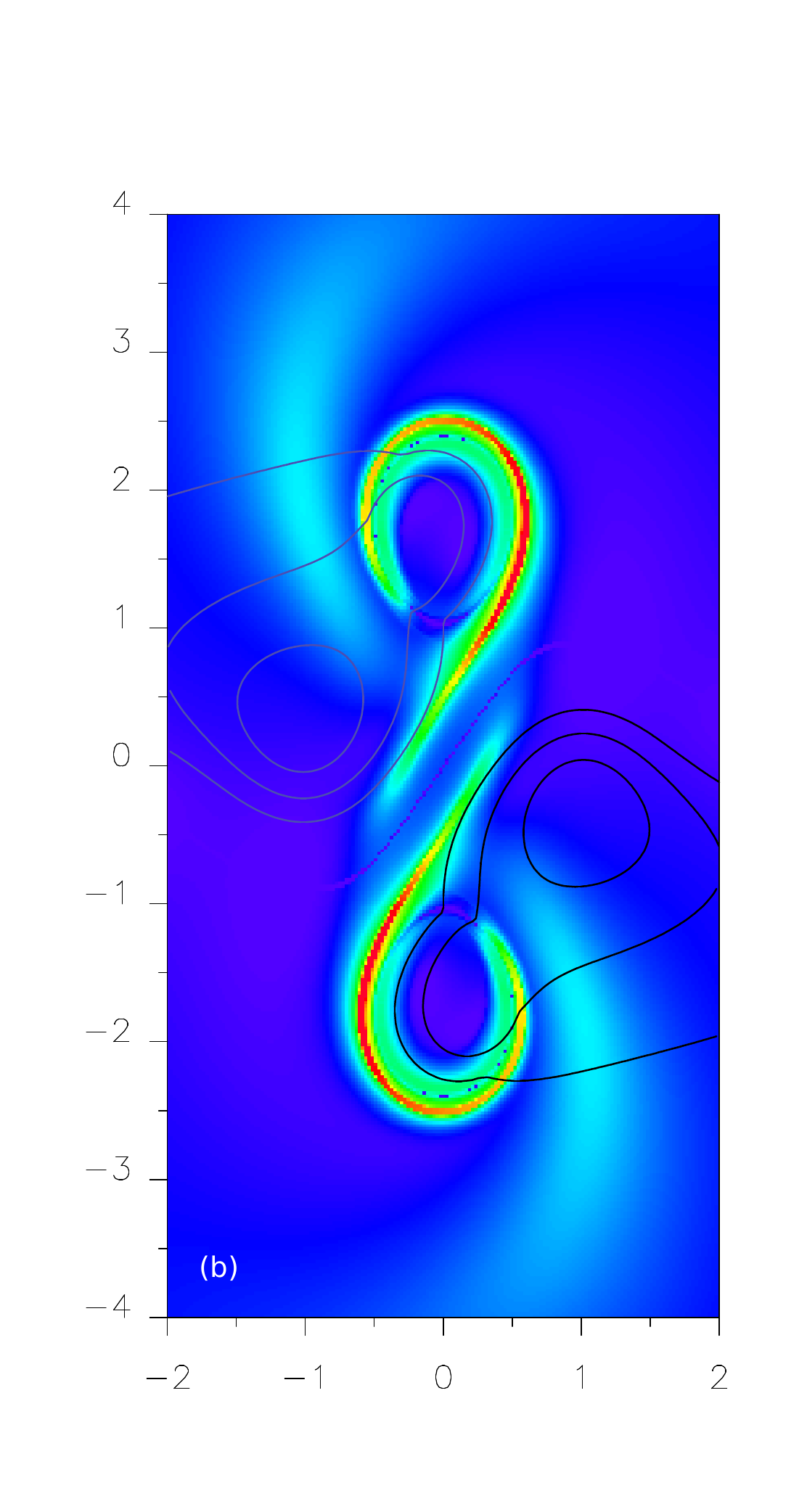}
  \includegraphics[scale=0.52, angle=0, trim = 0.5cm 0.0cm 1.0cm 0.0cm, clip]{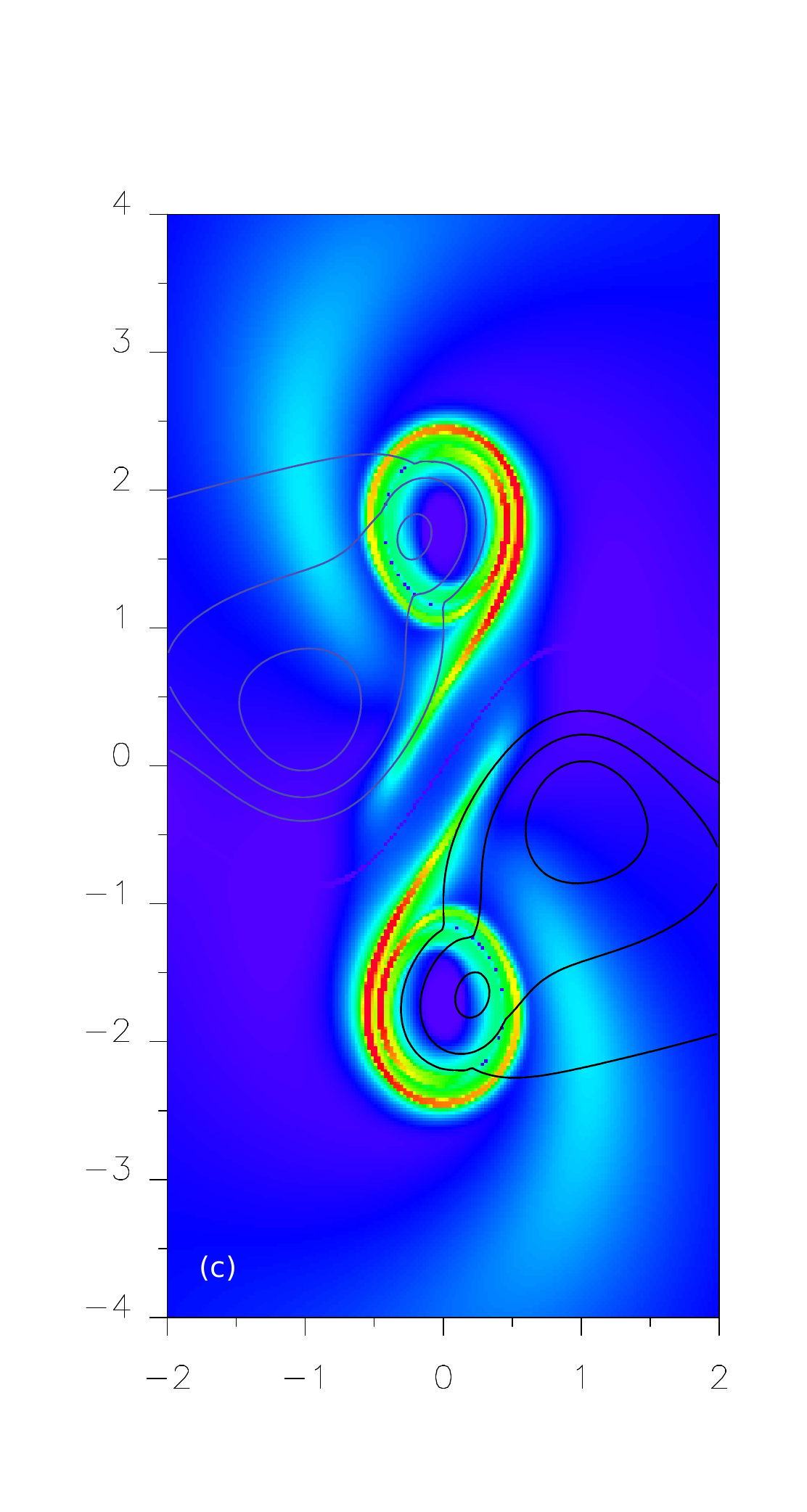}
  \caption{Photospheric QSL foot prints of the Titov-D\'emoulin flux
           rope model \citep{1999A&A...351..707T}  for different values
           of the flux-rope twist.
           The colour map shows the logarithm of the magnitude of the squashing factor
           $Q$. The QSLs are the region in red and yellow
           where $Q$ is large. Panels (a) and (b) shows the hook-shaped QSLs
           of moderately twisted ropes with 1 and 1.5 turns respectively.
           Panel (c) shows the spiral-shaped QSLs of a highly twisted rope
           with a twist of 2 turns. The black and gray lines are contours of the
           normal component of the magnetic field.}

  \label{ep_fig_qsls}

\end{figure}

\begin{figure}
\centering

  \includegraphics[scale=0.45,trim = 0.5cm 0.5cm 3.2cm 0.0cm, clip]{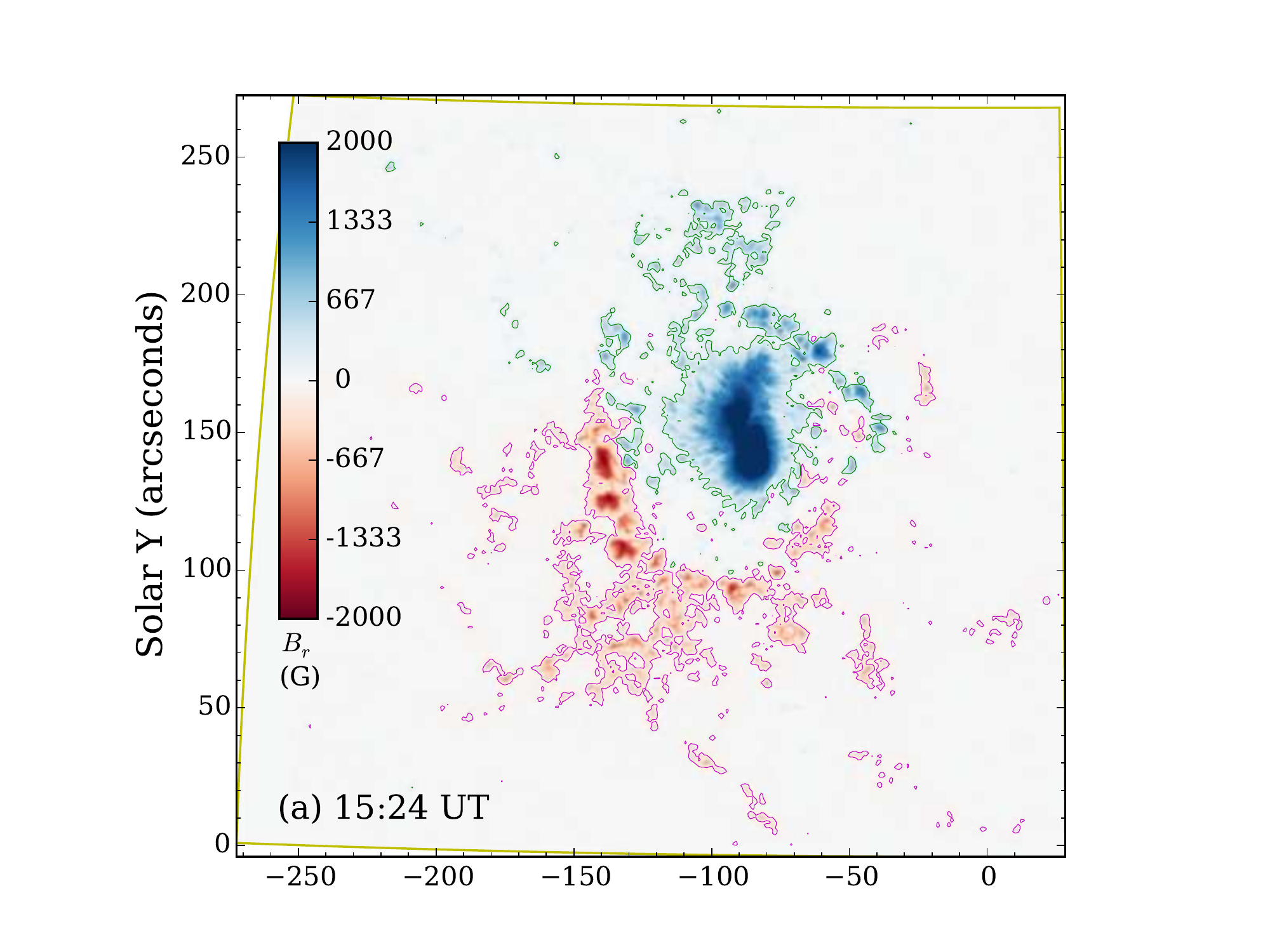}
  \includegraphics[scale=0.45,trim = 2.5cm 0.5cm 3.2cm 0.0cm, clip]{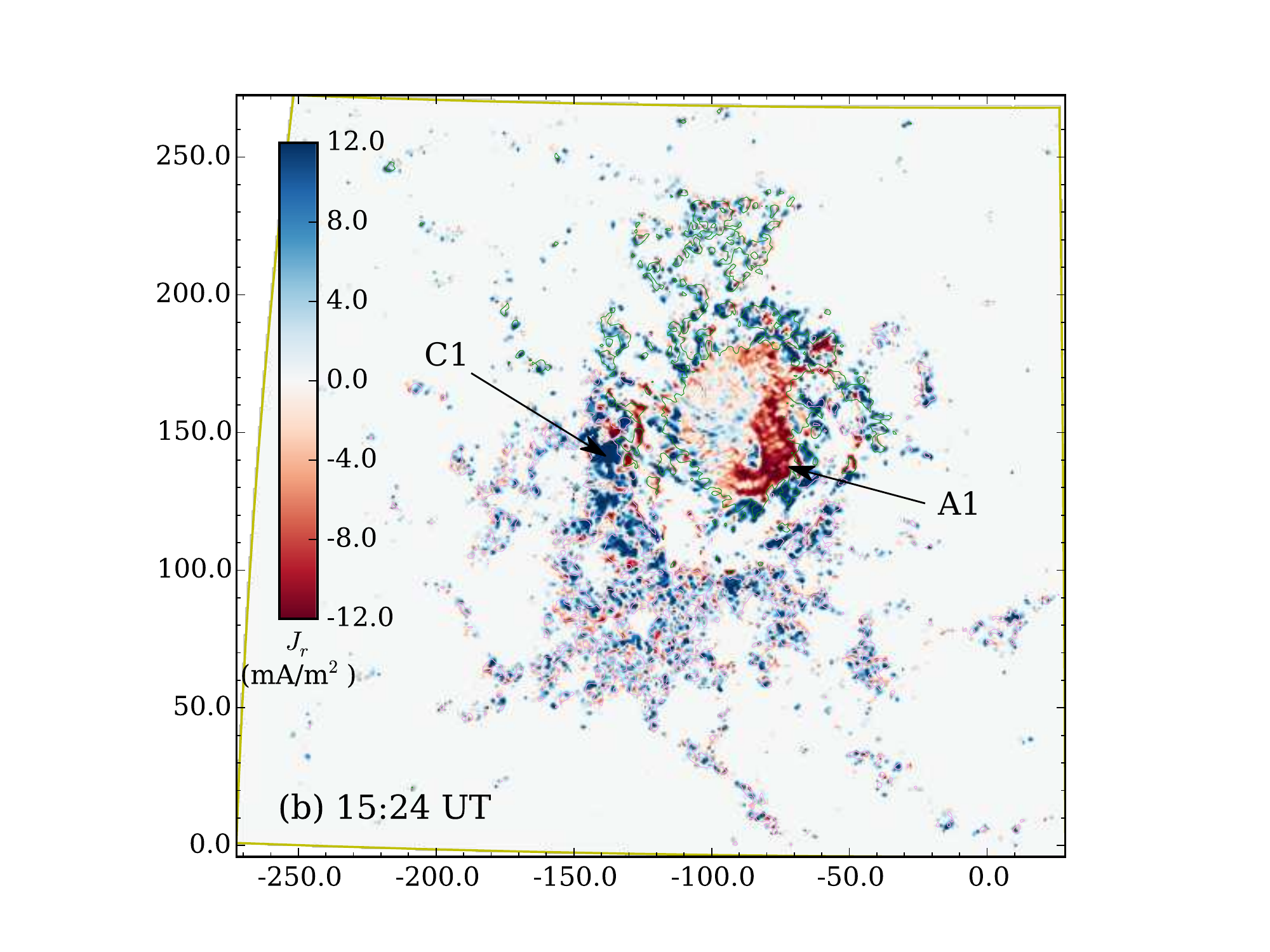}\\

  \includegraphics[scale=0.45,trim = 0.5cm 0.0cm 3.2cm 1.0cm, clip]{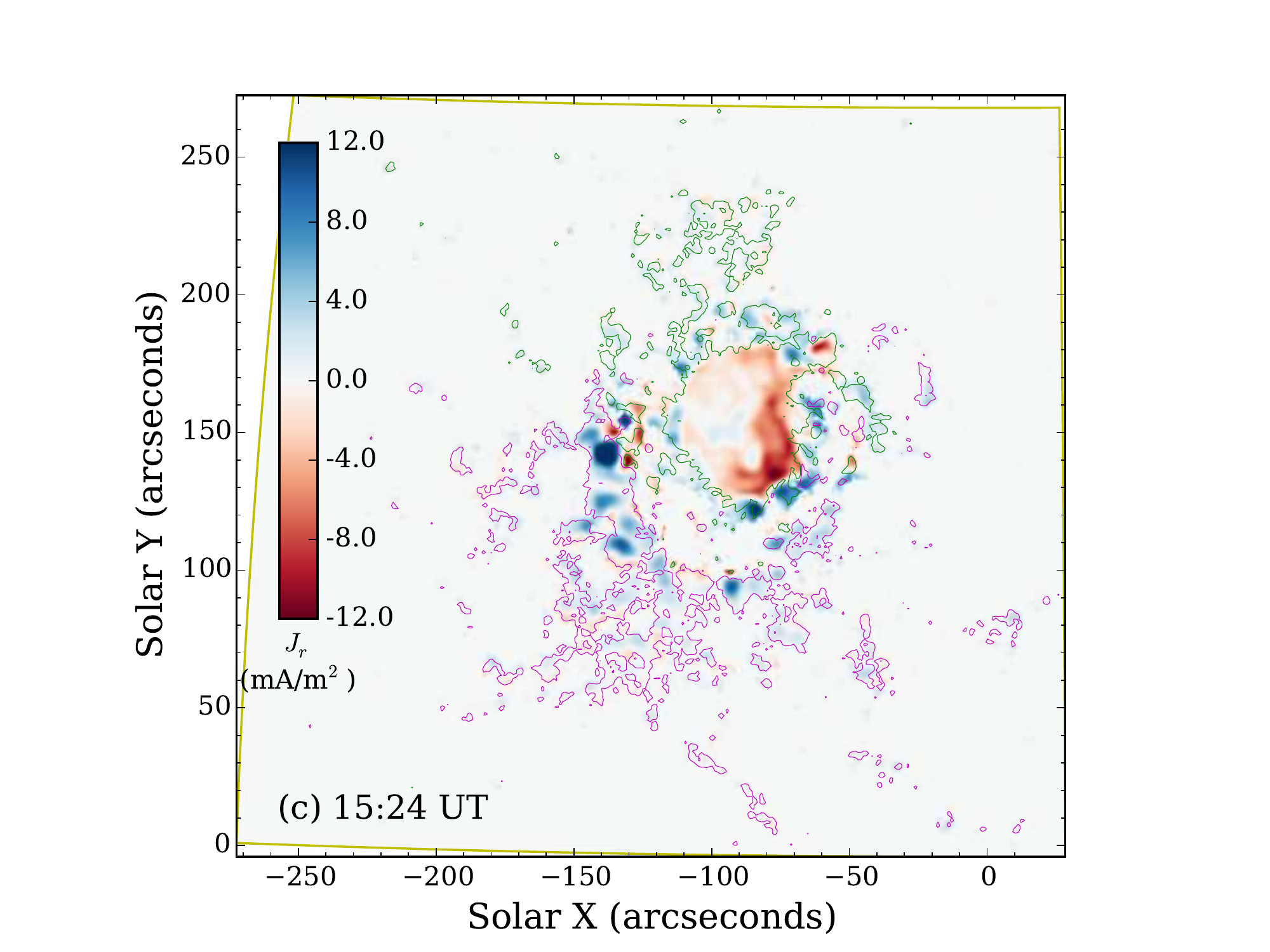}
  \includegraphics[scale=0.45,trim = 2.5cm 0.0cm 3.2cm 1.0cm, clip]{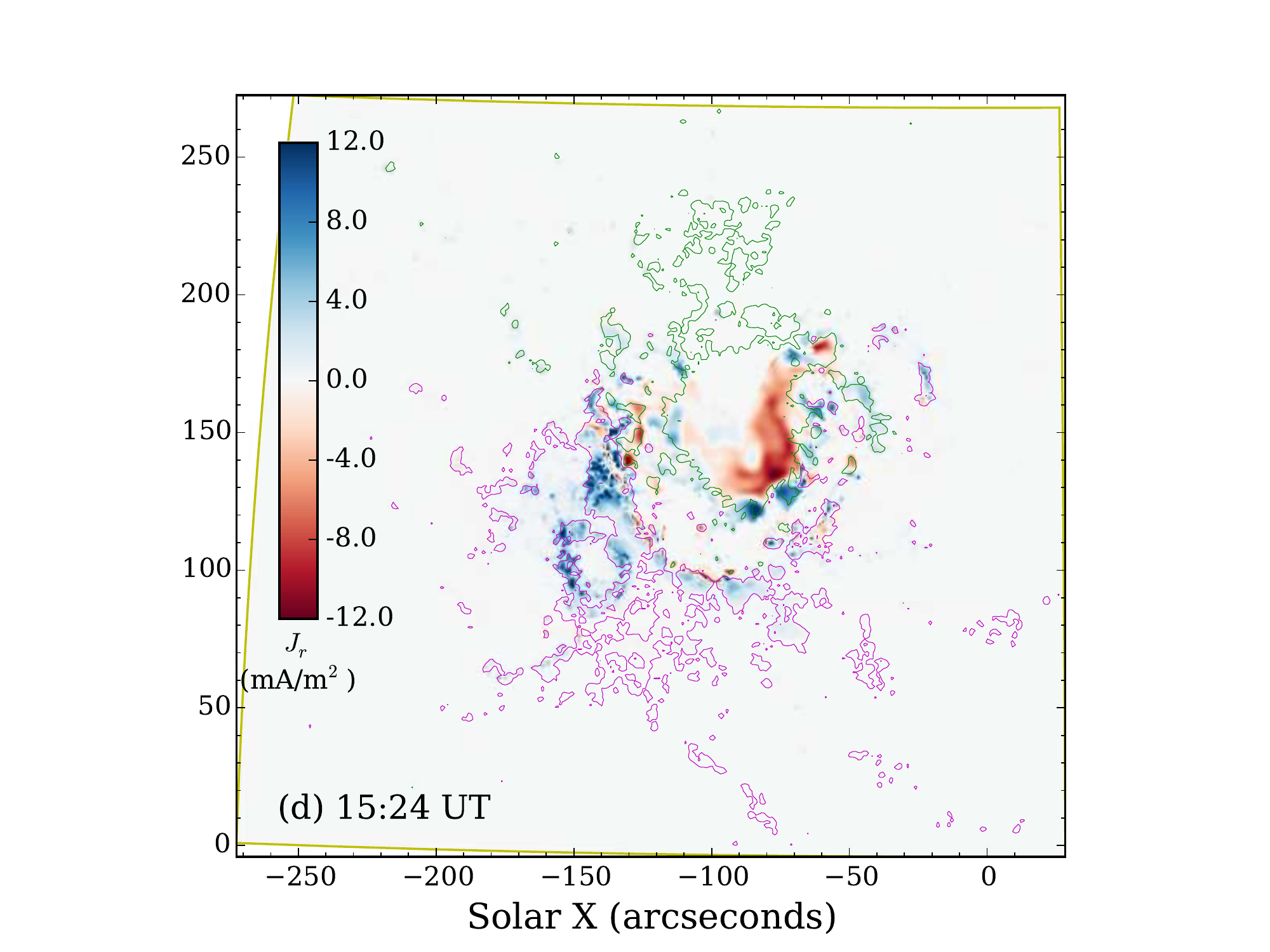}
  \caption{The photospheric boundary conditions for the NLFFF model of AR 12158.
           Panel (a) shows $B_r$. Panel (b) shows $J_r$ computed from
           the HMI data without smoothing. Panel (c) shows $J_r$ computed
           from the HMI data after first smoothing the transverse component
           of the magnetic field at the photosphere. This $J_r$ is used
           to compute the $\alpha_0$ boundary conditions for the force-free
           model. Panel (d) shows
           distribution of $J_r$ from $P$ solution of the nonlinear
           force-free model. Where $B_r<0$ (inside the purple contours)
           the distribution of $J_r$ is computed from the model rather than being fixed as a boundary
           condition (see Section \ref{section_extrapolation}). The
           purple and green lines are contours of $B_r$ at $\pm 200\, {\rm G}$ respectively. }

\label{sg_fig1}
\end{figure}

\begin{figure}

\centering

  \includegraphics[scale=0.35]{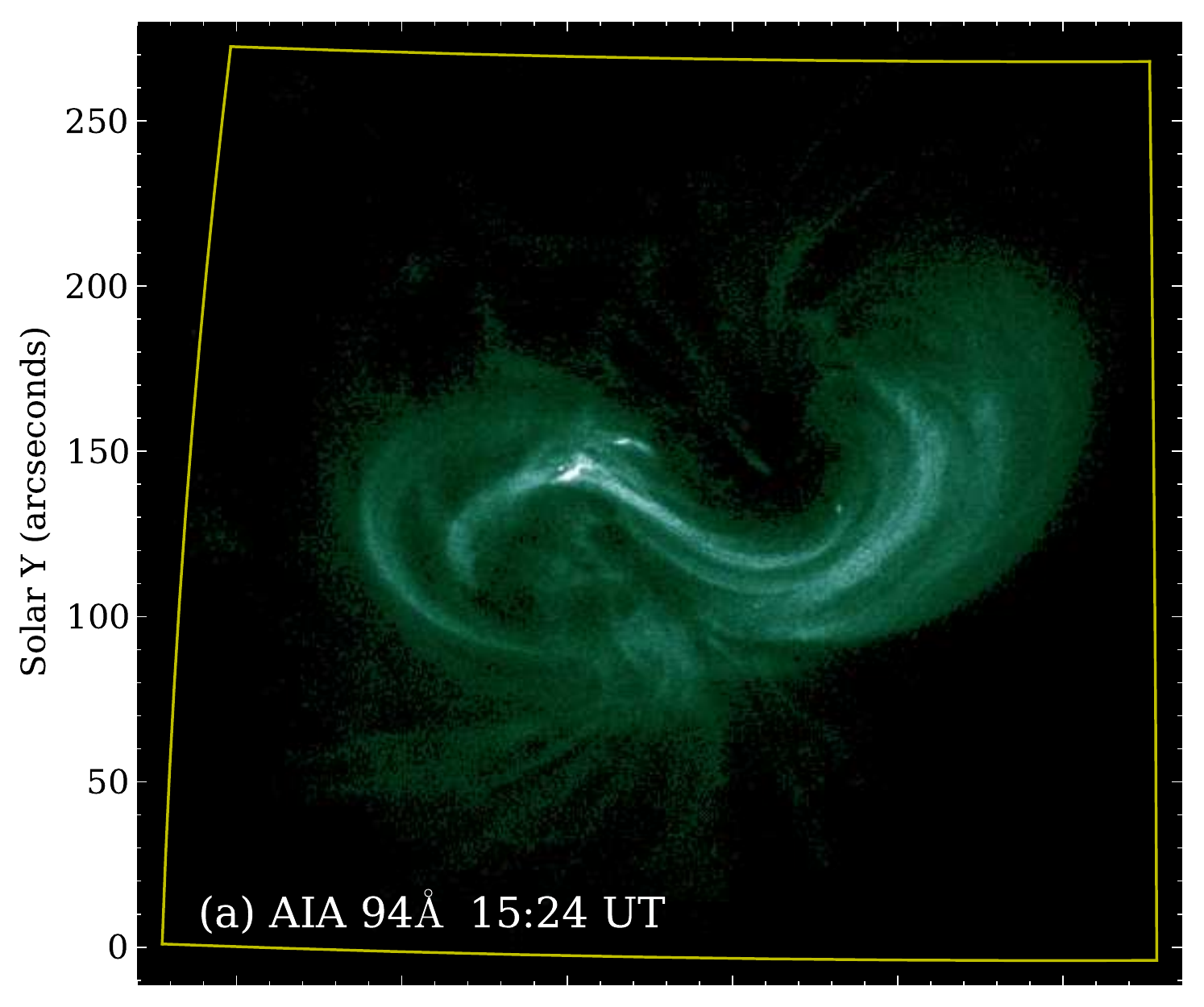}
  \includegraphics[scale=0.35]{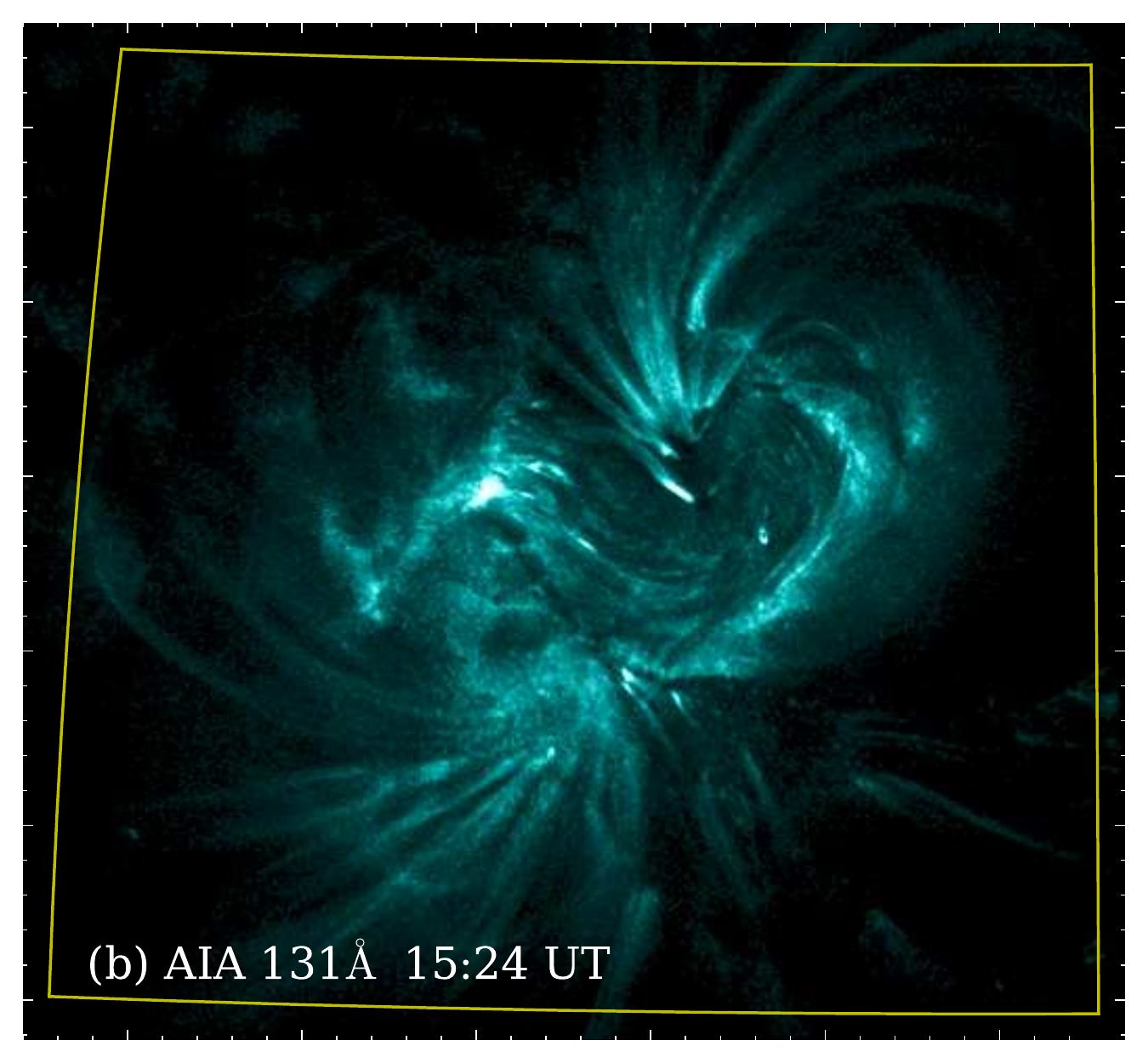}
  \includegraphics[scale=0.35]{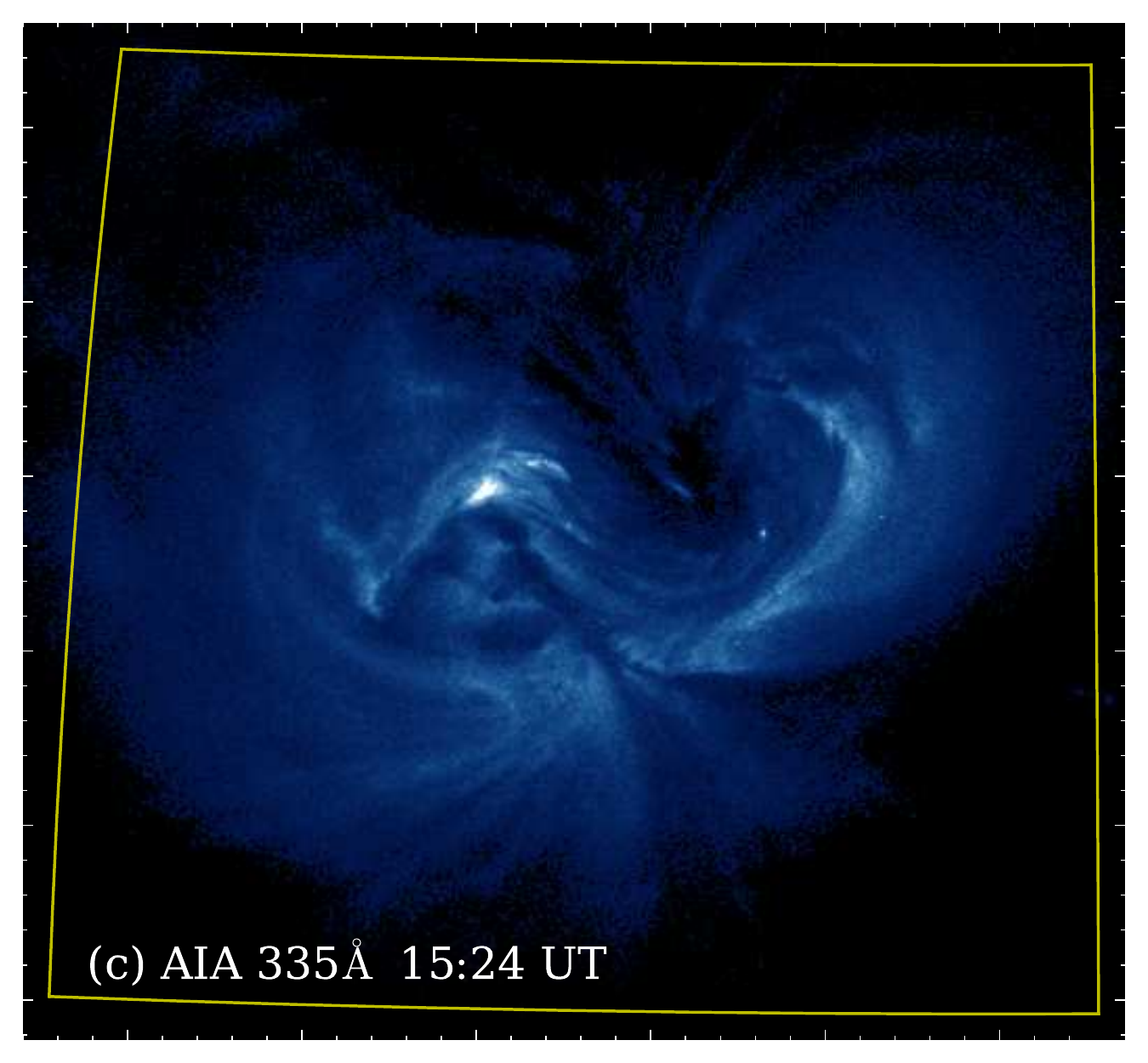}\\

  \includegraphics[scale=0.35]{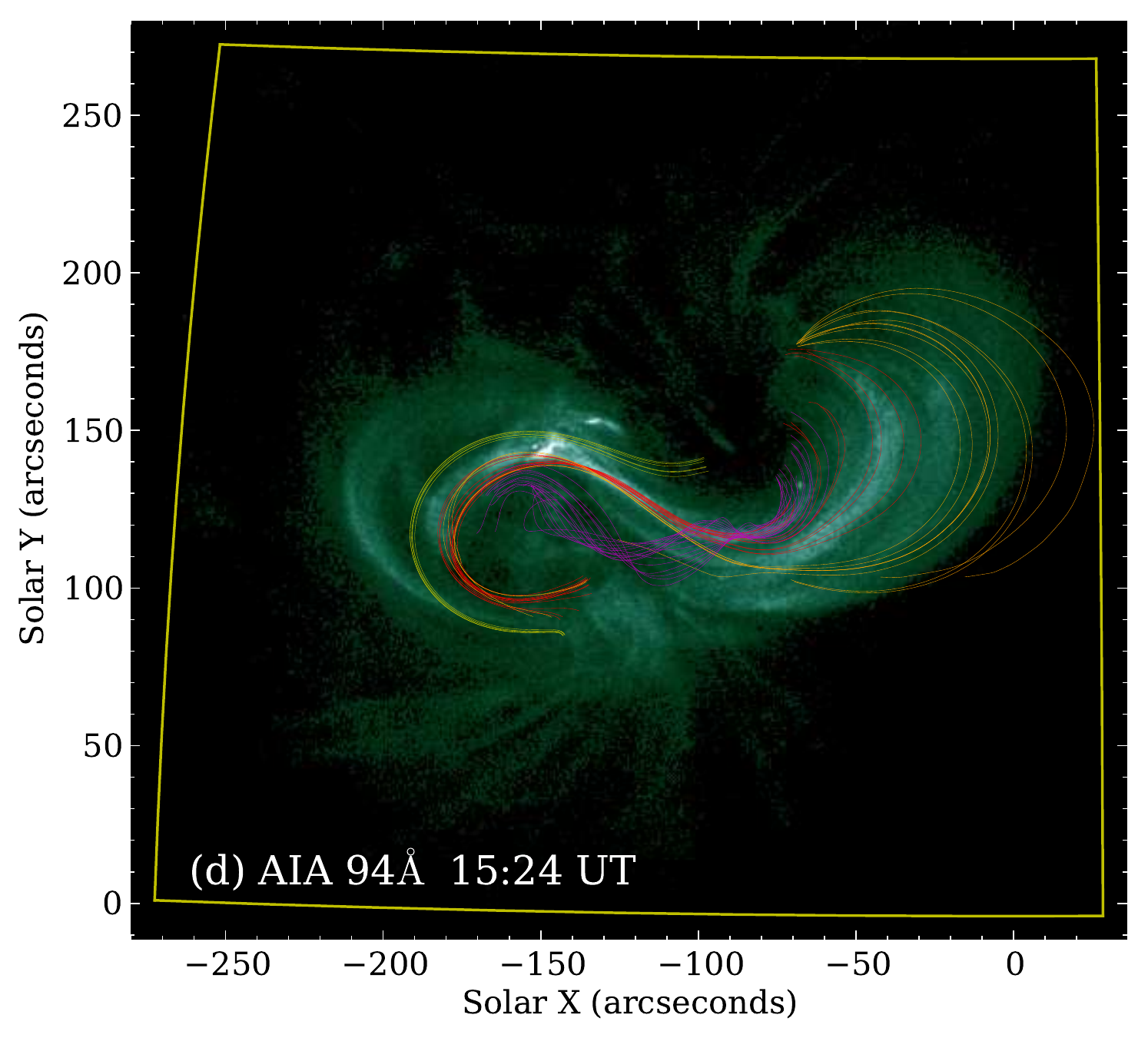}
  \includegraphics[scale=0.35]{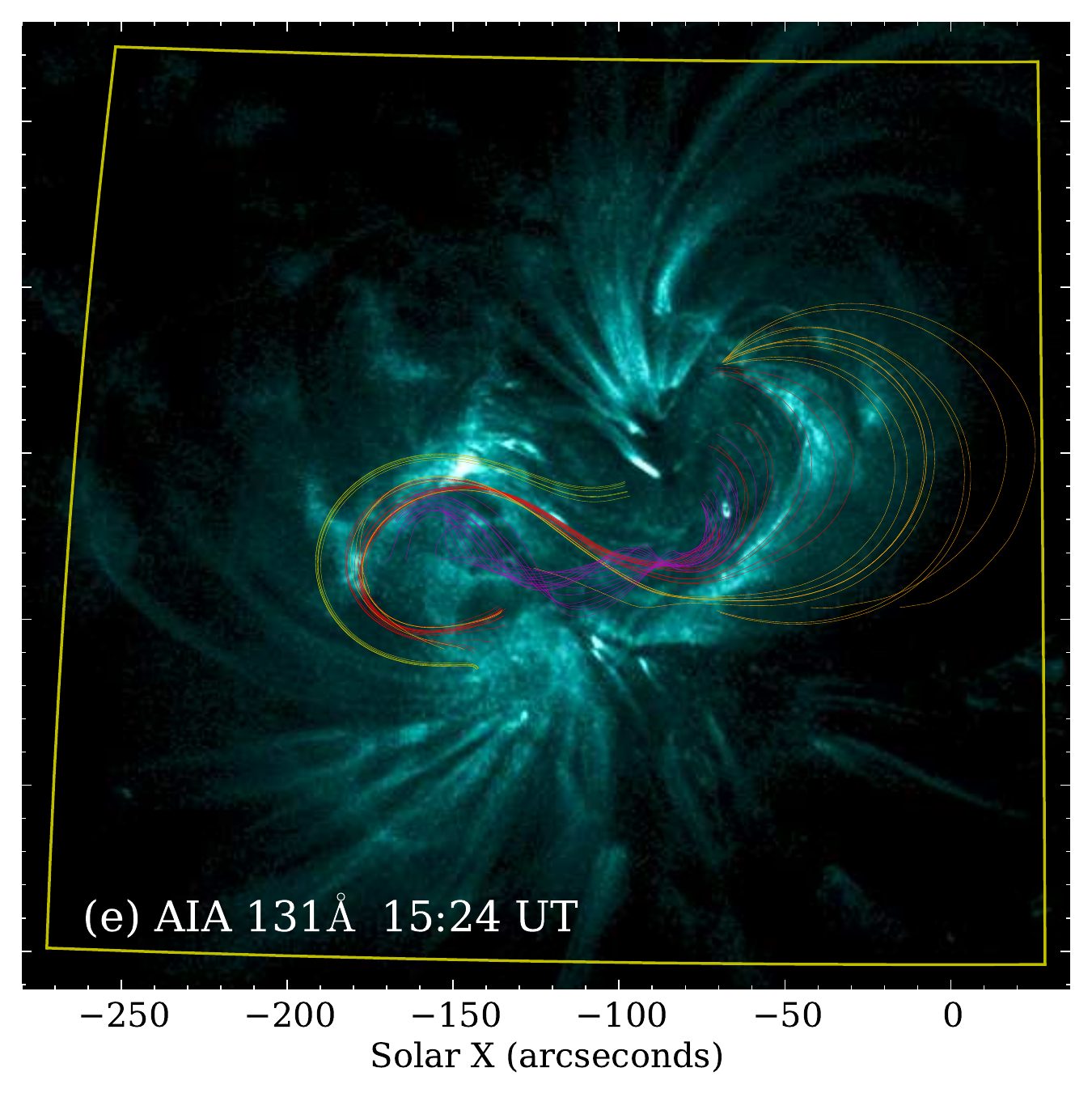}
  \includegraphics[scale=0.35]{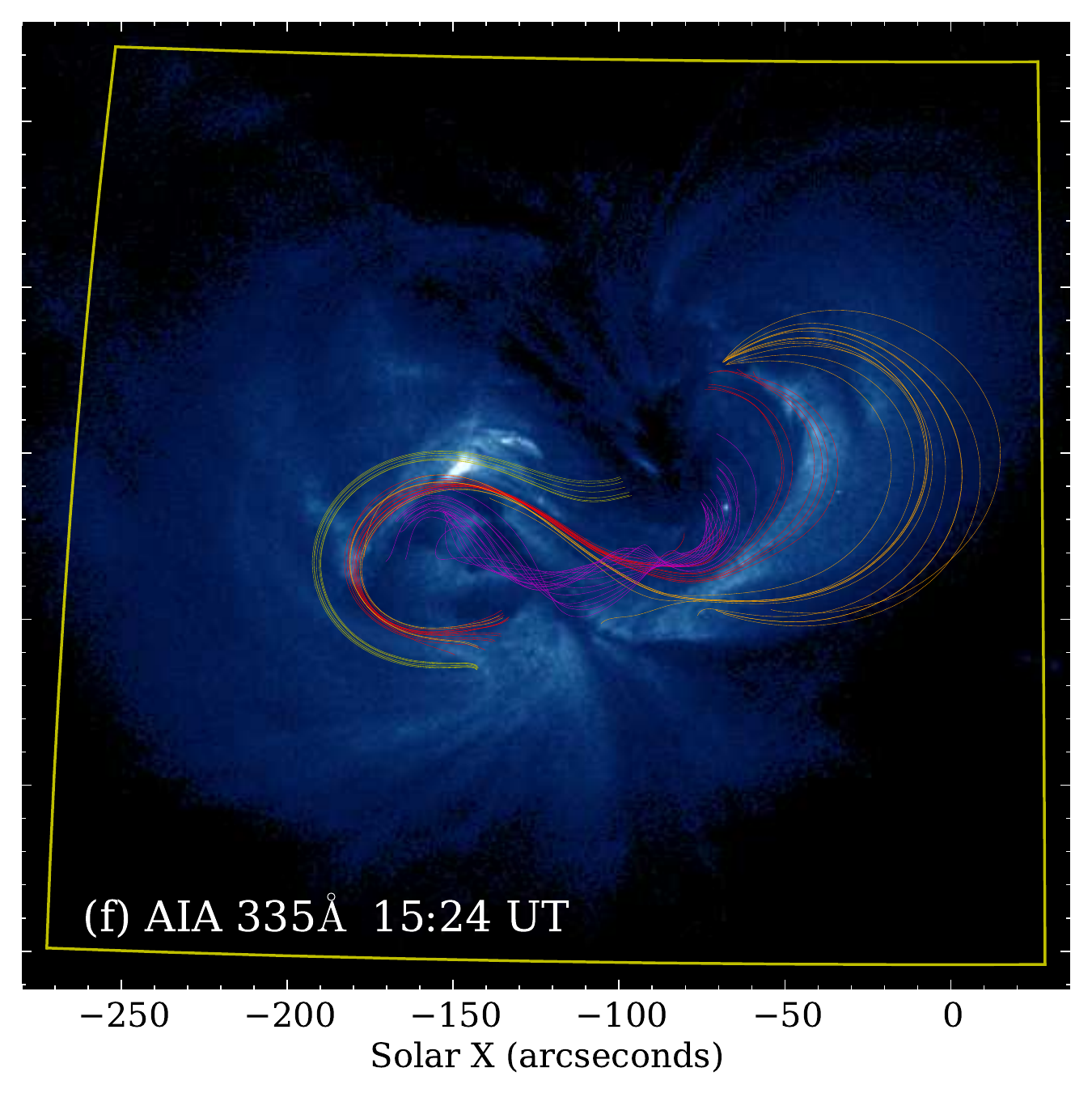}\\

  \includegraphics[scale=0.39]{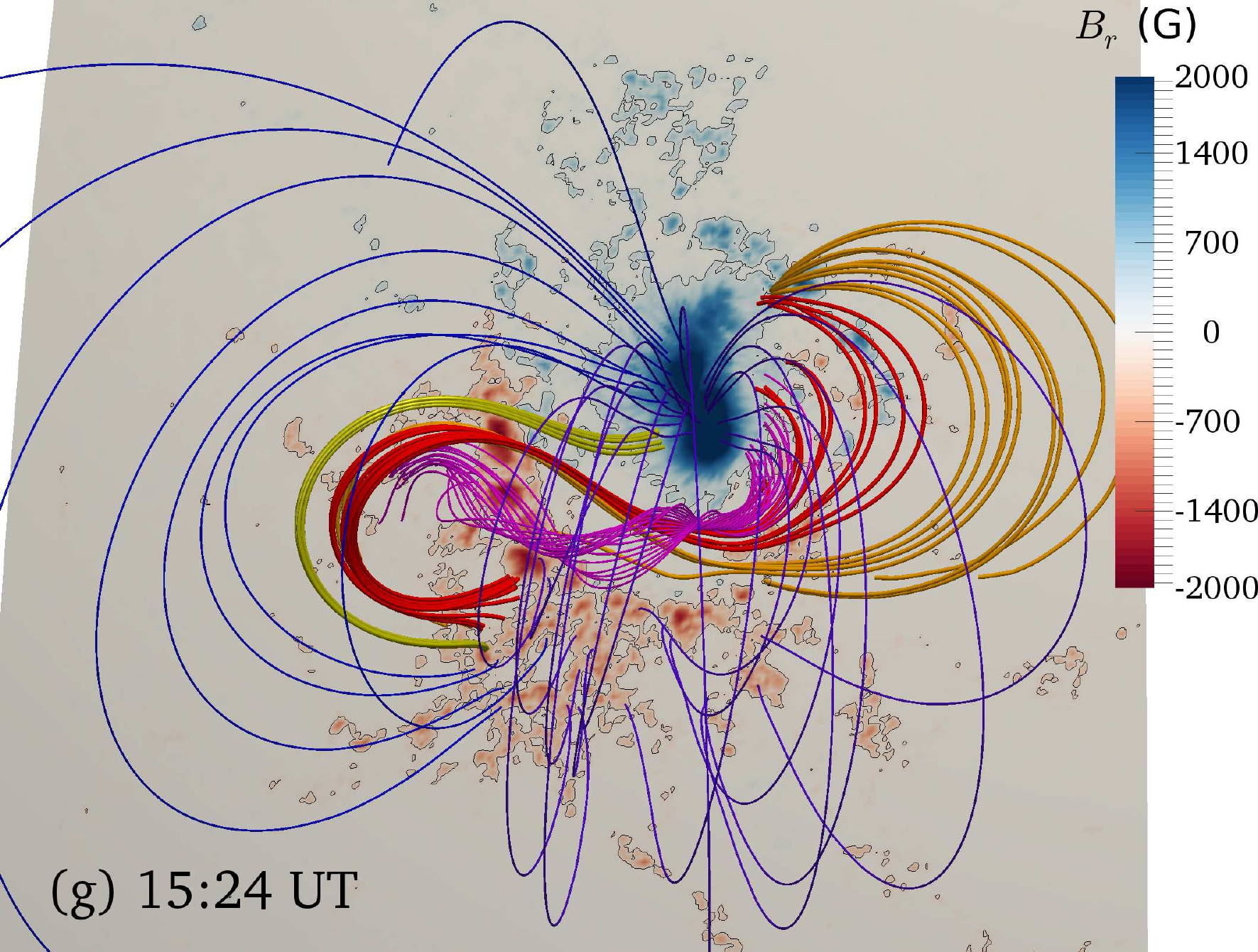}
  \includegraphics[scale=0.39]{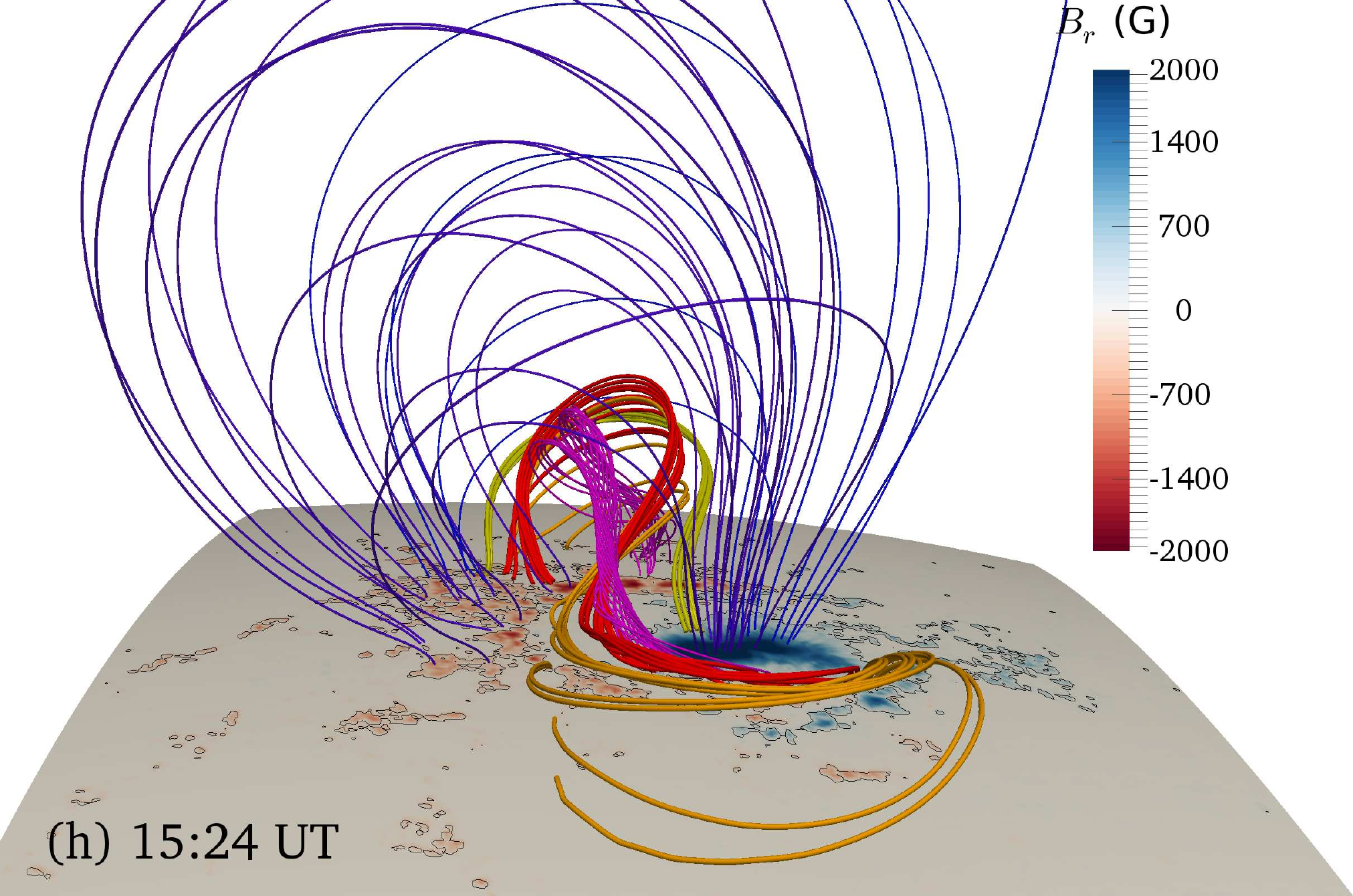}

  \caption{The coronal plasma of AR 12158 seen in EUV along
           with the field lines of the extrapolation.
           Panels (a)-(c) are AIA images of the region at 94\AA\,
           131\AA\, and 335\AA, respectively. Panels (d)-(f) are the
           same AIA images with selected field lines from the force-free
           model superimposed. Panels (g)-(h) are three-dimensional views of
           extrapolation field lines. Except for the blue field lines, which correspond to overlying arcades
           the field lines drawn in panels (d)-(e) are the same
           as those in panels (g)-(h). Panel (g) is a view of the region from the direction
            of Earth, and panel (h) is a view of the region looking
           towards solar east. The photosphere is coloured to show
           $B_r$, and the black lines are $B_r=\pm200$G contours.
           The yellow region in panels (a)-(f) shows the photospheric
           extent of the calculation volume. The purple flux rope field has no
 observational counterpart in the AIA  channels and in XRT image. The red, yellow and orange field lines trace the sigmoid.}

  \label{sg_fig0}
\end{figure}

\begin{figure}
\centering
\includegraphics[scale=0.9]{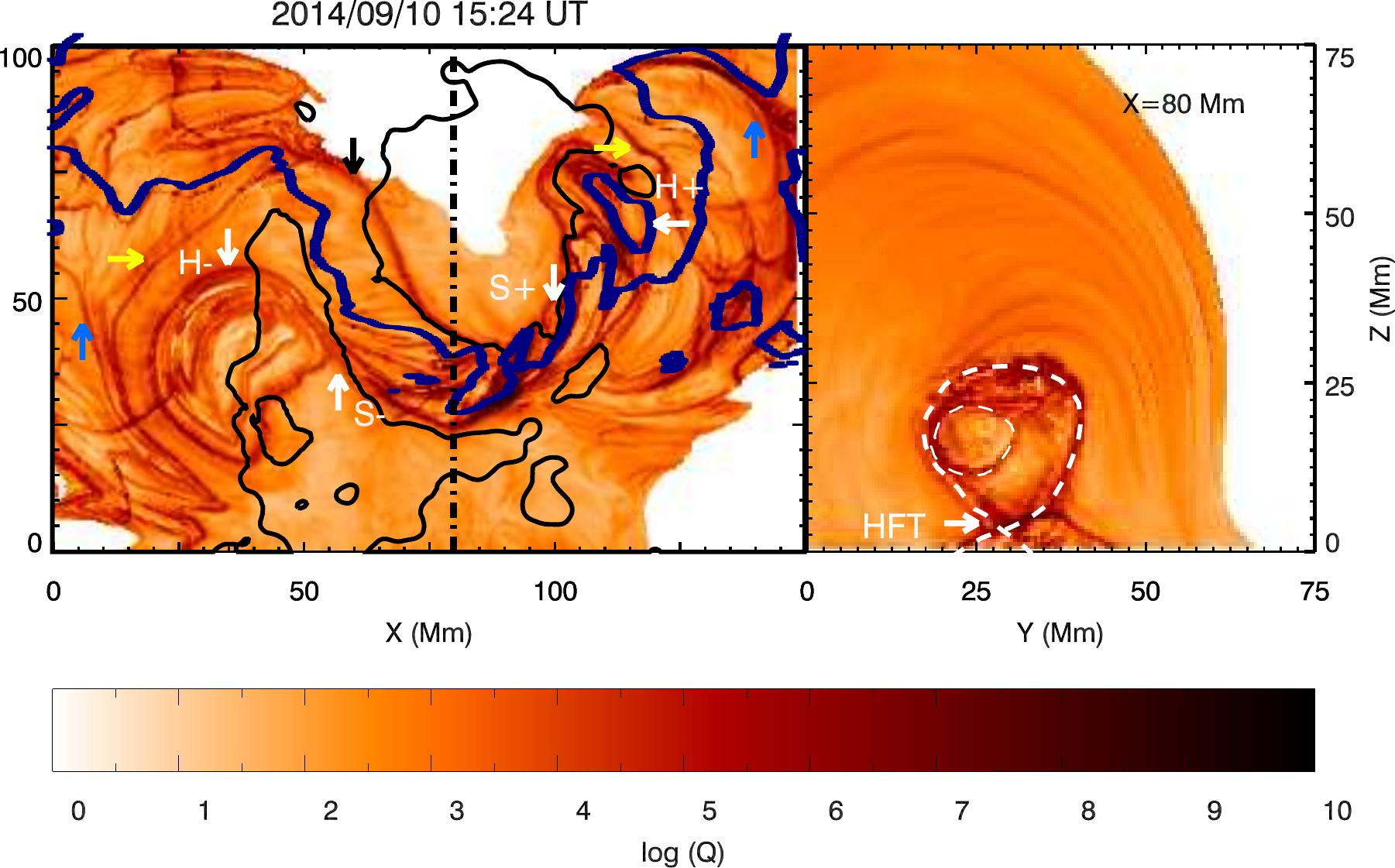}
\caption{The left panel shows the distribution of the squashing factor $Q$
at the photosphere at 15:24 UT (almost two hours before the X1.6 flare).
The locations with relative high $Q$ are the QSLs.
The black curves are the $\pm 200\, {\rm G}$ contours of $B_z$,
and the blue curve is the polarity inversion line.
The white arrows show the four parts (H+, S+, H-, S-)
of the double inverse-J shaped QSLs. The yellow arrows show the external
QSL-hooks and the black arrow displays an elongated QSLs in the north
of the PIL. The blue arrows indicate QSLs which separate the AR magnetic field with its environment.
The vertical dot-dashed line shows the
location of the slice cut shown in the right panel, where X equal
80 Mm.  In the right panel, the dashed white line  shows the inverse-tear drop-shaped envelope of the flux rope and the core of the flux rope in its middle, the white arrow indicates the location of the HFT structure.}

\label{jz_fig0}
\end{figure}

The left panel of Figure \ref{jz_fig0} shows map of $Q$ at the lower boundary.
The dot-dashed black line indicates the location of the vertical plane in which $Q$ is computed
and the corresponding map is displayed on the right panel of Figure \ref{jz_fig0}. At the center of the horizontal plot,
there are pronounced double inverse-J shaped QSLs
that correspond to the main body of the flux rope, which may consist of different substructures like the different QSLs (the
red , yellow, orange field lines in Figure \ref{sg_fig0}).
The straight sections (S+ and S-) coincide with the
PIL, and the hook sections (H+ and H-) are located on either side of the
PIL (H+ is located in the positive polarity and H- in the negative one).
It is interesting to compare these QSLs to those
derived from the TD flux rope shown {\bf in}
Figure \ref{ep_fig_qsls}. The QSLs in Figure \ref{jz_fig0}
are hook shaped, but do not {\bf completely encircle} the
leg of the rope as in the panel (c) of Figure \ref{ep_fig_qsls}.
It would appear that the flux rope obtain by the NLFFF
extrapolation has a twist between 1 and 2 turns. This is comparable
to the number of turns seen in Figure \ref{sg_fig1}.
The yellow arrows indicate hooked QSLs at the periphery that
are parallel to the H+ and H- hooks. These are related to the field
lines in yellow and orange in Figure \ref{sg_fig0}.
There are more external QSLs shown by the blue arrows. The QSLs represent the boundary
between non sigmoidal active region connections and long remote field lines.
They are not associated with the sigmoid.


The complex shape of the hooked QSLs with multiple arcs is reminiscent of a complex
distribution of twist, more complex than in all the idealized models quoted in the Section
\ref{section_introduction}, except for the magneto-frictional
ones \citep{Savcheva2012} that are also complex.

The black arrow indicates another elongated QSL to the north of
the PIL that encircles the positive polarity sunspot and stretches to the east.
This QSL separates the field
lines within the positive polarity into two systems. The southern
system is a sheared arcade with relative low height in the vicinity of
the flux rope (see the yellow field lines in Figure \ref{sg_fig0}, panels (d) - (h)).
The northern system overlays the flux rope
(see the blue field lines in Figure \ref{sg_fig0}, panels (g) and (h)).

We found the cross section of an inverse-drop QSLs volume (white dash line in heavy) in the HFT
(indicated by the white arrow) in the right column of Figure \ref{jz_fig0}. This inverse-drop QSLs
defines the different quasi-connectivity between the flux rope and
the overlaying arcades (the blue field lines in Figure \ref{sg_fig0},  panels (g) and (h)).
The height of the crossing section at HFT is relatively low and is considered to be
the preferential place for reconnection \citep{2005A&A...444..961A,
2013A&A...555A..77J,Savcheva2012,2012ApJ...744...78S,2015ApJ...810...96S}.

The inverse-drop QSLs volume is relatively large, extending up to 25 Mm in height and with a similar width. This is in agreement
with the size of the teardrop structure found in \citet{Savcheva2012} and an order of magnitude larger than the one described in
\citet{2014ApJ...787...88Z}. This is another confirmation of the presence of a mature/well developed flux rope in the system.
This teardrop topological structure is also embedded in larger weaker surrounding QSLs as previously described \citep{Savcheva2012,2013ApJ...779..157G}.
Another circular QSL (white dash line in light) is embedded inside the inverse-drop QSL,
which demonstrates the different quasi-connectivity inside the flux rope,
and confirms the complex distribution of twist in the flux rope.
It is interesting to examine the correspondence between the field lines in
Figure \ref{sg_fig0} and the QSLs in Figure \ref{jz_fig0}.
We find that the flux rope in Figure \ref{sg_fig0} (purple field lines)
crosses the slice at Y $\sim$ 33 Mm and Z $\sim$ 25 Mm and does not pass through the
small circle. The field lines that pass through the circle are longer and have a sigmoid shape similar to the
orange field lines shown in Figure \ref{sg_fig0}.


%
\subsection{Comparison with Flare ribbons in the rising phase of the flare}
\label{section_flare_ribbons}

\begin{figure}

\centering
\includegraphics[scale=1.1]{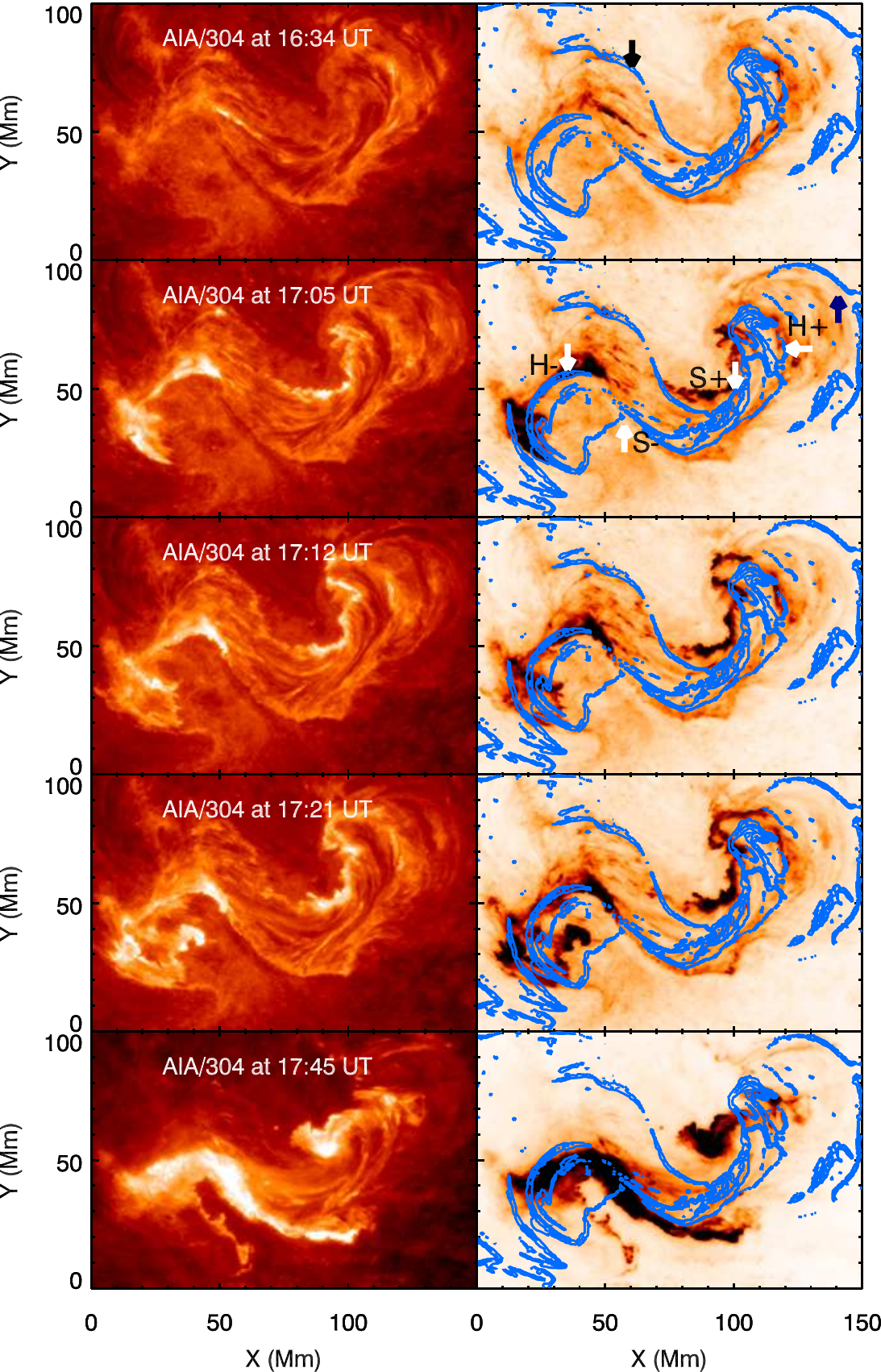}
\caption{Flare ribbons as observed in 304 $\AA$ by AIA before and up to
the peak of the flare (left and right columns, the latter in reverse color), and
QSL footprints calculated from the single NLFFF extrapolation of the HMI magnetogram
at 15:24 UT (right column, overplotted on all the 304 $\AA$ images).
The arrows in the first two rows of the right column show the same structures of the QSLs
being displayed on Figure \ref{jz_fig0}.
}
\label{jz_fig1}
\end{figure}

Figure \ref{jz_fig1} shows a comparison between the QSL footprints
at the lower boundary and the observed flare ribbons.
The left column consists of AIA 304\AA\ images showing the flare ribbons
at different times during the eruption. The right column consists
of the same AIA images with the QSL footprints superimposed.

The top row of Figure \ref{jz_fig1} shows brightenings develop
well before the peak of the GOES X-ray flux at 17:45 UT.
They have a similar shape as the double inverse-J shaped QSLs.
Besides the brightening along these QSLs, there is also a faint one at
the north-east along the elongated QSLs, at $(X,Y) \approx$ $(50\,{\rm Mm}, 80\,{\rm Mm})$,
the location is indicated by the black arrow in the top right panel.
The right panels show the QSLs superimposed on the AIA image and
demonstrate that the faint elongated brightening structure is the continuation
(within the same polarity) of the intense, compact ribbon that forms at the north-west of the
big sunspot later. The strong correlation between the brightening and the QSLs
indicates that reconnection is occurring in the corona and the energy is transported
along the 3D QSLs towards the footprints depicted in Figure \ref{jz_fig1} before the flare.
Repeated brightenings are often observed \citep[][]{2015ApJ...804...82C}
and are interpreted as indicating that successive reconnection happens well before
the impulsive phase of the flare.

While the GOES  1$-$8 \AA\ soft X-ray light curve indicates that the
flare begins at 17:21 UT, the UV observations shows that the
evolution of the ribbons begins earlier. Between 17:05 and 17:21 UT,
the hooked parts show zipping brightenings and their position evolves
noticeably. Thus, the ribbons are very dynamic before the official
start time of the flare at 17:21 UT.
There is a relatively good agreement between the QSL footprints and the flare ribbons
when they first appear at 17:05 UT. Some special features of the QSLs
are labeled by arrows in the right column. The labels have the same meaning
as those in Figure \ref{jz_fig0}. We note an excellent match between
the location and shape of the positive hook (H+) and
the straight part of the negative ribbon (S-). Although the
straight part of the positive ribbon (S+) is a few Mm north of the QSL,
there is good agreement between the shapes. The
negative hook (H-) also appears slightly shifted relative to the HFT footprint
but, again, there is a good morphological agreement.

We expect that the two hours of difference between the time of the extrapolation/topological
analysis and the actual time of the flare is responsible for this small shift in location. We speculate that during that time the
currents within the flux rope have grown and that the footprint of flux rope occupies a larger area, in particular in the more diffuse positive polarity.
This can easily explain the difference of position between the QSL and H+.

To our knowledge, the present analysis shows the best match ever published between the position of the hooks of J-shaped flare ribbons
and the topological analysis. With their extrapolation, \citet[][cf. Figure 10]{2014ApJ...787...88Z} were
not able to capture the correct J shape of the
observed ribbons.  As discussed in \citet{2015ApJ...810...96S}, the flux-rope insertion method was used to match seven different two J-shaped flares,
the largest sample so far. They could recover well the straight part of the ribbons in all cases
except for one involving a B-class flare. However, they were
able to fit the hooked parts of the ribbons with the QSLs in only half of
the studied regions and with an overall morphological agreement lower
than in the present study. Since the flux-rope insertion method does
not use vector magnetogram data, it is not as strongly constrained as the
Grad-Rubin method. In addition, the magneto-frictional approach used in  \citet{2015ApJ...810...96S} involves an ad hoc positioning of the extremities
of a flux rope and the code is free to relax the foot points of the flux rope independently of the actual electric current concentrations.
The excellent morphological agreement observed here between the position of the QSLs obtained solely from the extrapolation of magnetic
information with fully independent UV observation at the onset of the flare,
provides a clear demonstration of the quality of the extrapolation.

Between 17:05 and 17:21 UT, the brightenings at the hooks evolve rapidly.
At 17:12 UT, the S+ ribbons moves further north and the H+ hook grows in size moving northward as well. On the contrary, the H- hook part appears to move inward towards the straight part S-. This tendency continues until 17:21 UT. Later during the flare, between 17:21 UT and the peak phase of the flare at 17:45 UT, the hook parts H+ and H- appears to shrink, while stronger emissions originates from the straight parts S+ and S-.

While there was a good agreement between the QSLs from the NLFFF model and the
flare ribbons when they first appear, the match becomes progressively worse
as the eruption progresses, and the flare ribbons change their shapes and locations.
In the framework of the CSHKP model in 2D, and of the recent standard model in 3D
that results from MHD \citep{2012A&A...543A.110A,2015SoPh..tmp...63J} and magneto-frictional
\citep{Savcheva2016} simulations, these changes in the ribbons merely
occur because of the ongoing flare reconnection during the eruption. Therefore it is natural that our
static force-free model, which was constructed two hours before the eruption, no longer
accurately represents the topology and therefore the ribbons on the long run.
Thus, the hook parts of the sigmoidal structure in 304\AA\ at the
start of the eruption, and before the peak of the GOES flare,
are recovered by our extrapolation because at this time
the coronal magnetic field had not changed significantly from its
force-free pre-flare state.




%
%
%
%
%
%
\section{Discussion}
\label{section_discussion}

We perform a NLFFF extrapolation of AR 12158 before a major
eruption with the goal of testing the predictions of an eruptive flare model
firstly proposed by \citet{1996JGR...101.7631D}, then presented in a series
of recent papers \citep{2012A&A...543A.110A,
2013A&A...549A..66A,2013A&A...555A..77J} and reviewed in \citet{2015SoPh..tmp...63J}.

The authors refer to their model as ``the standard flare model
in three dimensions'', which will also do in the following discussion.
This model makes specific predictions regarding the topology of the
pre-eruption magnetic field, which we test by
performing a topological analysis of our extrapolated magnetic field.

The reliability of the topological analysis depends fundamentally
on the reliability of the force-free extrapolation, so it is important
to be mindful of the limitations of this kind of modeling.
Generally, extrapolation methods perform better when applied to
idealized test cases than to vector magnetogram data, which means that the results rely
not only on the reliability of the extrapolation methods but also on the quality of the vector magnetograms
(see the discussion in \citealt{2009ApJ...696.1780D}).

A particular problem is that generally force-free methods do not
achieve a self-consistent solution when applied to solar data,
i.e. the extrapolation contains significant residual Lorentz forces.
This is a problem for all methods, see for example the comparison
of methods by \citet{2009ApJ...696.1780D} and \citet{2015ApJ...811..107D}.
One specific advantage of the Grad-Rubin method, however,
is that it is based on a well-posed formulation of the force-free
boundary value problem and tends to satisfy the solenoidal condition
($\nabla \cdot \mathbf B=0$) even when applied to solar data,
while other methods may not \citep{2009ApJ...696.1780D}.

The force-free extrapolation of AR 12158 encountered similar problems.
We do not achieve a self-consistent solution because the Grad-Rubin
did not strictly converge. We found that we could improve the convergence of our method
either by excessive smoothing of the boundary, or by setting $k$ to
a small value. However, these solutions did not agree
with the AIA images. The close agreement between the
field lines of the less smoothed solutions and the AIA images indicates that the
extrapolation likely reproduces the topology of the magnetic field.
This is because the Grad-Rubin method does not specify a priori
the connectivity of the field.
For this study, this is more important than achieving
a strictly force-free magnetic field by excessive smoothing of the
boundary data. Nevertheless, it is important
that the results presented here are considered with these caveats in mind.

The force-free model contains a clear magnetic flux rope as illustrated
by the purple field lines in Figure \ref{sg_fig0}. The appearance of
the flux rope is similar to those obtained during flux-cancellation
MHD simulations of bipolar regions \citep{2012A&A...543A.110A} --
the overlying arcade, the twisted core, and the highly twisted S-shaped
sigmoid appear in both cases.

We compute the squashing factor $Q$ for the NLFFF extrapolation
at the photosphere and a cross-section of the flux rope (see Section \ref{section_topology}).
The double inverse-J shaped QSLs at the photosphere, the inverse-drop QSLs
and the HFT structure on the vertical plane found in our observation have been
seen in both analytic models \citep{1999A&A...351..707T} and
MHD calculations (e.g., \citealt{2012A&A...543A.110A}).
The appearance of an HFT in our extrapolation, which is non-idealized and based on real solar data,
supports ``the standard flare model in three dimensions'' as a realistic model.

One interesting point from our extrapolation is the complex shape of
the hooked QSLs, which wrap around the legs of the flux rope. Firstly
they have multiple arcs which are almost parallel to each other
at both ends of the flux rope. These features may be related with
the multi-threaded curved ends of the sigmoid.
Secondly, a QSL with a nearly circular cross-section
is embedded inside the inverse tear drop shaped coronal QSL. These
features all together tend to indicate that the twist distribution within the
flux rope is much more complex in the extrapolation than in idealized models. The
shapes of the inner hooked QSLs (H+ and H-) at the lower boundary (that are strongly
curved and have a spiral-like shape), and those of the extrapolated field lines, both consistently
show that the core of the flux rope in AR 12158 is more twisted than its envelope,
and has between 1 and 2 turns.

We compare the distribution of $Q$ on a plane $\approx 1.5\,{\rm Mm}$
above the photosphere to the flare ribbons.
The QSLs on the plane closely match the flare
ribbons at the start of the eruption. We note that the hook-shaped (J-shaped)
extremities of the flare ribbons are recovered by the model and
coincide with a QSL that wraps around the legs of the flux rope.
This is consistent with the predictions of ``the standard flare model in three dimensions'' \citep{2015SoPh..tmp...63J}.
The nonlinear force-free model is an equilibrium state,
so it cannot be used to study the flare from its peak phase onwards.

%
%
%
%
%
%

\section{\gui{Summary and} conclusion}
\label{section_conclusion}

In this paper we
successfully checked some predictions of ``the standard flare model in three
dimensions'' \citep[see][]{2015SoPh..tmp...63J} : the occurrence
of double J-shaped QSL footprints below sigmoids ; their match with observed
flare ribbons ; and their link with a coronal HFT.
This result was obtained by performing
a NLFFF extrapolation and a topological analysis
of AR 12158, two hours before a sigmoid eruption and an
X-class flare that took place on September 10, 2014.

The extrapolation was performed with
the recently developed CFITS force-free modeling code in spherical
geometry, that uses the Grad-Rubin method
\citep{2014SoPh..289.1153G}. Although the NLFFF model
was calculated in spherical coordinates, we performed
the topological analysis
assuming a flat Cartesian geometry, by calculating the squashing factor $Q$ with the QSL
code developed by \citet{2014ApJ...787...88Z}. This geometrical
approximation has been used by others (e.g.\ \citealt{2015ApJ...810...96S})
who found that it only introduces minor errors for compact
regions, such as AR 12158.

Among the various extrapolations that we produced, by varying several parameters
of the method, we found one that produced a relatively good match
between the sigmoid as observed with AIA
and the modeled field lines.
The extrapolated coronal field contains a magnetic flux rope, of which
handedness also matches the observed inverse-S shape
of the sigmoid. The distribution of
the squashing factor $Q$ at a plane $\approx 1.5\,{\rm Mm}$ above the photosphere
revealed a double inverse J-shaped QSL footprint located on both sides of the polarity inversion line
below the sigmoid. This is characteristic of coronal flux ropes. The complex hook-shaped
extremities of the QSL footprint display a spiral-like shape. This is reminiscent of
idealized models of flux ropes of which core has slightly more than one turn. This
is also consistent both with the field lines of modeled coronal flux rope itself,
and with the distribution of $Q$ as calculated in a coronal plane orthogonal both
to the photosphere and to the flux rope axis, that reveals a closed quasi
circular-shaped QSL embedded inside a larger flux-rope related inverse tear-drop
shaped QSL, below which an HFT in present.

Several other observed features that are well reproduced by our extrapolation
are worth noticing. Firstly, its
vertical current density distribution at the photosphere resembles that derived from the HMI
instrument, in particular in the plage flux
concentrations where {\bf $J_r$} was
not prescribed as boundary
conditions for the calculation. Secondly, we found
a relatively good agreement between the QSLs of the model and the flare ribbons observed by
AIA early in the flare, before its EUV/SXR peak, when the coronal field can still be
represented by the pre-flare NLFFF. In particular,
a surprisingly good match was achieved
in the shape and location of the hook-shaped ribbons and QSL.
Through its resulting QSL and HFT, our
topological analysis straightforwardly explains why
slip-running reconnection \citep{Aulanier2007}
occurred in this flaring active region, as previously reported by
\citet{2015ApJ...804L...8L} and \citet{dud2015}

These agreements
between the NLFFF model and the two independent sets of observations (AIA and HMI)
firstly and fully confirm the aforementioned
predictions of the ``new standard flare model in 3D''
\citep[as compiled in][]{2015SoPh..tmp...63J}.
They are also consistent with results recently obtained with a different NLFFF code based
on the magnetofrictional method \citep{2015ApJ...810...96S,Savcheva2016}.
Secondly, they confirm the capacity of the Grad-Rubin method in general,
in modeling solar sigmoids accurately, and in recovering thick coronal flux ropes
\citep[as previously achieved by][]{2004A&A...425..345R,2010ApJ...715.1566C}.
And thirdly, they support the quality of the CFITS code and of the methodology
used in this paper, in particular, in their joint capacity in identifying key flux rope properties,
namely (i) their strong-to-weak twist distribution from its core to its outer parts,
(ii) the location and extent of their photospheric endpoints, both of which being
hard to obtain with other NLFFF methods, and (iii) the existence of a coronal HFT
below the flux rope at the flare onset, that must lead to slip-running reconnection.

Following \citet{2015ApJ...810...96S}, our results also support the claim that flare
ribbons constitute one of the most complete sets of information that can be obtained on the
geometry of erupting structures and on the sites of flare energy
release, which are all together important elements of space weather studies.
In this context, NLFFF models that can accurately reproduce the location and the shape of flare ribbons
should henceforth result in reliable pre-eruptive
flux rope properties, such as their magnetic twist,
field strength, orientation, and free energy, as well as the location of its underlaying HFT. Such models may then
be used not only to predict the location of the flare energy release through slip-running
reconnection at the HFT, but also as initial conditions for MHD simulations of CME
initiation and further propagation towards Earth.


\acknowledgements
The data have been used by courtesy of NASA/SDO and the HMI science team.
SDO is a mission of NASA's Living With a Star program. We would like to
thank P. D{\'e}moulin, J. Dud{\'i}k and Y. Liu for fruitful discussions
as well as the referee for helping us clarifying several points that
were initially ambiguous.
J. Zhao and H. Li are supported by the National Basic Research Program of China
under grant 2011CB811402, by NSFC under grant 11273065, and by the
Strategic Pioneer Program on Space Sciences, Chinese Academy of Sciences,
under grant XDA04076101. J. Zhao is also supported by NSFC under grant 11503089, 11522328 and 11473070.
Support for S.A. Gilchrist at LESIA was provided by a contract from
the DIM ACAV and R\'{e}gion
Ile-de-France. And his support at NWRA is provided by the NASA Living With
a Star programthrough grant NNX14AD42G, and by the Solar Terrestrial program
of the National ScienceFoundation through grant AGS-1127327.


\bibliographystyle{apj}
\bibliography{zhao2015_revised_references} 

%
%
%
%

%
%
%
%

%
%
%
%
%
%

%
%
%

%
%
%

\end{document}